\documentclass{mn2e}

\usepackage{graphicx, psfig, times, amsfonts}

\newcommand{\HP}{{\sc hipparcos}}

\newcommand{\M}{{\sc 2mass}}

\begin{document}

\title[Improved Baade-Wesselink surface-brightness relations]{Improved Baade-Wesselink surface-brightness relations
%\thanks{Table~1 is available in electronic form.}
}

\author[Martin A.T. Groenewegen]{
M.A.T. Groenewegen \\ 
Instituut voor Sterrenkunde, PACS-ICC, Celestijnenlaan 200B, 
B--3001 Leuven, Belgium
}

\date{received: 2004,  accepted: 2004}

\maketitle
\begin{abstract}
Recent, and older accurate, data on (limb-darkened) angular diameters
is compiled for 221 stars, as well as $BVRIJK[12][25]$ magnitudes for
those objects, when available. Nine stars (all M-giants or
supergiants) showing excess in the $[12-25]$ colour are excluded in
the analysis as this may indicate the present of dust influencing the
optical and near-infrared colours as well. Based on this large sample,
Baade-Wesselink surface-brightness (SB) relations are presented for
dwarfs, giants, supergiants and dwarfs in the optical and
near-infrared.  M-giants are found to follow different SB-relations
from non-M giants, in particular in $V$-$(V-R)$.  The preferred
relation for non-M giants are compared to earlier relation by Fouqu\'e
\& Gieren (1997, based on 10 stars) and Nordgren et al. (2002, based
on 57 stars). Increasing the sample size does not lead to a lower rms
value. It is shown that the residuals do not correlate with
metallicity at a significant level.  The finally adopted observed
angular diameters are compared to those predicted by Cohen et
al. (1999) for 45 stars in common, and there is reasonable overall, to
good agreement when $\theta <$6 mas. Finally, I comment on the common
practice in the literature to average, and then fix, the zero point of
the $V$-$(V-K)$, $V$-$(V-R)$ and $K$-$(J-K)$ relations, and then
re-derive the slopes. Such a common zero point at zero colour is not
expected from model atmospheres for the $(V-R)$ colour and depends on
gravity. Relations derived in this way may be biased.
\end{abstract}


\section{Introduction}

The $PL$-relation of Cepheids is of fundamental importance in
establishing the cosmic distance scale.  Determining whether or not
the slope of the Galactic relation is different from that for the LMC
and/or SMC Cepheids relations is of prime importance, as this could imply a
metallicity dependence of the slope of the relation, with important
consequences for the application of the $PL$-relation to other
galaxies.  The apparent zero point and slope of the fundamental (FU)
and first-overtone (FO) $PL$-relations in the LMC and SMC have now
been well determined in $V, I, W$ (the reddening free, so-called
Wesenheit-index based on $V$ and $I$), and $J$ and $K$ (Udalski et
al. 1999, Groenewegen 2000, Nikolaev et al. 2004, Sandage et al. 2004). 

The situation is less clear for our Galaxy. The accuracy of the
\HP\ parallax data was not high enough to determine slope {\it and}
zero point. Instead, the data had to be analysed in a statistical way,
such that for an {\it assumed} slope, a zero point could be derived
(e.g., Feast \& Catchpole 1997; Groenewegen \& Oudmaijer 2000)

There are basically two alternatives to obtain direct distance
estimates to Galactic Cepheids, namely, using Cepheids in open
clusters (e.g. Feast 1999, Tammann et al. 2003) where the distance to
the cluster has been obtained in another way (basically main-sequence
fitting), or, in combining linear diameter determinations (as
determined from integration of the radial velocity curve and assuming
a projection factor), with angular diameter determinations coming from
direct measurements using an interferometer (e.g. Kervella et al. 2003) 
or coming from a surface-brightness (SB) relation and a
reddening-corrected magnitude (e.g. Fouqu\'e et al. 2003, Storm et al. 2004).

For the latter method to work one has therefore to determine and
calibrate accurate SB relations. There is a whole body of work on this
subject (e.g., Barnes et al. 1978; Di Benedetto 1993 [DB93]), and more
recently, with direct application to Cepheids, the works by Fouqu\'e
\& Gieren (1997; herafter FG97), and Nordgren et al. (2002; hereafter
Nord02). The latter calibration is used by Fouqu\'e et al. (2003) to
establish the most recent slopes and zero points of the Galactic
$PL$-relation for FU Cepheids in $BVIWJHK$ colours.
 
In addition, SB-relations are important in estimating angular
diameters for planning interferometric observations (either to check
if a potential science target would be resolved, or to check if a
calibrator is indeed unresolved, for a given baseline), and for
distance estimates in eclipsing binaries (Salaris \& Groenewegen 2002).

The calibration by Nord02 is based on 57 giants observed with the Navy
Prototype Optical Interferometer (NPOI) by this group (Nordgren et
al. 1999; Nordgren et al. 2001). On the other hand there have been
other recent papers presenting new diameter determinations (e.g., 69
stars in van Belle et al. 1999), and there exist older data of high
quality; recently, Richichi \& Percheron (2002) conveniently presented
a catalog of high angular resolution measurements.

The aim of the present paper is to present updated SB relations based
on the largest set of accurate angular diameter determinations. 
Section~2 presents the angular diameter data and the search for
apparent magnitudes. Section~3 briefly recalls the relevant
equations, and Section~4 presents the results. The discussion in
Section~5 ends this paper.

\begin{table*}
%\footnotesize
\setlength{\tabcolsep}{1.0mm}
\caption{Identification (HR or IRC identifier, unless otherwise
listed in the remarks), adopted angular diameter and error, adopted
visual reddening and error, $BVRIJK$[12][25] photometry, reference to
the angular diameter and photometry (when not taken from SIMBAD as
discussed in the text), and remarks, of the sample studied. An entry
with $-9.99$ means that this magnitude is not available.
%The complete table available in electronic form.
}
\begin{tabular}{rlrcrcrrrrrrrrll} \hline

ID   & Spectral Type & $\theta$ & $\sigma_{\theta}$ & $A_{\rm V}$ & $\sigma_{\rm A_V}$ & $B$ & $V$ & $R$ & $I$ & $J$ & $K$ & $[12]$ & $[25]$ & References & Remarks \\ 
     &             & (mas) & (mas) &       & & & & & & & &  & \\ \hline

    21 & F2\_IV\_SB   &  2.12 &  0.05 &  0.04 &  0.02 &  2.61 &  2.27 &  1.96 &  1.76 &  1.65 &  1.48 &  1.36 &  1.30 & Nord99  &  \\  
    79 & K5\_III      &  2.51 &  0.05 &  0.23 &  0.14 &  7.39 &  5.79 &  4.44 &  3.44 & -9.99 &  1.80 &  1.64 &  1.59 & vB99, 6, 7 &  \\  
   163 & G8\_III      &  1.77 &  0.08 &  0.08 &  0.02 &  5.25 &  4.38 &  3.70 &  3.19 &  2.84 &  2.21 &  2.14 &  2.11 & Nord99    &  \\  
   165 & K3\_III\_SB  &  4.14 &  0.04 &  0.05 &  0.01 &  4.56 &  3.28 &  2.36 &  1.70 &  1.24 &  0.48 &  0.29 &  0.27 & MAH03   &  \\  
   168 & K0\_IIIa     &  5.61 &  0.06 &  0.11 &  0.03 &  3.40 &  2.23 &  1.45 &  0.85 &  0.42 & -0.25 & -0.49 & -0.39 & MAH03    &  \\  
   259 & M4\_III      &  3.67 &  0.10 &  0.27 &  0.01 &  7.83 &  6.20 &  5.33 &  3.73 &  2.21 &  1.03 &  0.86 &  0.75 & vB99, 14   &  \\  
   265 & G8\_IIIb     &  1.63 &  0.11 &  0.18 &  0.06 &  5.60 &  4.64 &  3.87 &  3.37 & -9.99 &  2.35 &  2.24 &  2.21 & Nord99    &  \\  
   274 & G6\_III      &  2.80 &  0.11 &  0.40 &  0.01 &  6.50 &  5.42 & -9.99 & -9.99 & -9.99 &  2.93 &  2.98 &  2.93 & vB99   &  \\  
   294 & G9\_III      &  1.74 &  0.10 &  0.05 &  0.04 &  5.24 &  4.28 &  3.50 &  2.98 &  2.60 &  2.00 &  1.90 &  1.78 & Nord99    &  \\  
   337 & M0\_III      & 13.81 &  0.13 &  0.06 &  0.02 &  3.62 &  2.05 &  0.81 & -0.19 & -0.81 & -1.83 & -2.11 & -9.99 & MJS91   &  \\  
   351 & G9\_III      &  1.64 &  0.07 &  0.14 &  0.01 &  5.69 &  4.66 &  3.90 &  3.36 & -9.99 &  2.36 &  2.30 &  2.32 & vB99   &  \\  
   389 & K5\_III      &  2.22 &  0.06 &  0.08 &  0.01 &  6.62 &  5.23 & -9.99 & -9.99 & -9.99 &  1.93 &  1.89 &  1.76 & vB99   &  \\  
   402 & K0\_III      &  2.08 &  0.03 &  0.10 &  0.03 &  4.65 &  3.59 &  2.83 &  2.27 &  1.84 &  1.18 &  1.06 &  0.98 & KTFS04  &  \\  
   424 & F7\_Ib-II    &  3.28 &  0.02 &  0.02 &  0.01 &  2.62 &  2.03 &  1.53 &  1.22 & -9.99 &  0.52 &  0.50 &  0.50 & Nord99, 3 &  \\  
   437 & G7\_IIa      &  2.64 &  0.11 &  0.08 &  0.06 &  4.59 &  3.62 &  2.90 &  2.40 &  2.03 &  1.43 & -9.99 & -9.99 & Nord99    &  \\  
   442 & G9\_IIIb     &  1.64 &  0.10 &  0.18 &  0.06 &  5.72 &  4.72 &  3.95 &  3.43 & -9.99 &  2.33 &  2.30 &  2.24 & Nord99   &  \\  
   450 & M2\_III      &  2.42 &  0.05 &  0.29 &  0.01 &  7.42 &  5.89 & -9.99 & -9.99 & -9.99 &  1.85 &  1.70 &  1.57 & vB99   &  \\  
   464 & K3\_III      &  3.76 &  0.07 &  0.08 &  0.01 &  4.85 &  3.57 &  2.61 &  1.96 &  1.53 &  0.79 &  0.68 &  0.64 & Nord99   &  \\  
   489 & K3\_IIIb     &  2.81 &  0.03 &  0.09 &  0.07 &  5.80 &  4.44 &  3.38 &  2.67 &  2.13 &  1.24 &  1.16 &  1.10 & Nord99   &  \\  
   564 & M2\_III      &  2.96 &  0.06 &  0.17 &  0.01 &  7.42 &  5.82 & -9.99 & -9.99 & -9.99 &  1.50 &  1.28 &  1.19 & vB99   &  \\  
   603 & K3\_IIb      &  7.84 &  0.07 &  0.27 &  0.16 &  3.31 &  2.10 &  1.16 &  0.48 &  0.02 & -0.81 & -0.95 & -1.01 & MAH03   &  \\  
   617 & K2\_III\_SB  &  6.88 &  0.04 &  0.00 &  0.01 &  3.15 &  2.00 &  1.16 &  0.54 &  0.10 & -0.64 & -0.70 & -0.83 & Nord99   &  \\  
   631 & M3\_III      &  3.75 &  0.11 &  0.37 &  0.01 &  7.35 &  5.70 &  4.83 &  3.49 &  2.09 &  0.93 &  0.69 &  0.63 &  vB99, 14  &  \\  
   643 & K4\_III      &  2.91 &  0.05 &  0.16 &  0.05 &  6.30 &  4.82 & -9.99 & -9.99 & -9.99 &  1.29 &  1.21 &  1.09 & MAH03   &  \\  
   681 & M7\_IIIe     & 48.20 &  0.60 &  0.10 &  0.07 &  4.46 &  3.04 &  1.03 & -0.87 & -1.96 & -3.07 & -5.19 & -5.94 & WDH00 & eliminated: $m_{12}-m_{25}$ \\
   736 & K5\_III      &  2.60 &  0.11 &  0.12 &  0.07 &  6.62 &  5.15 & -9.99 & -9.99 & -9.99 &  1.66 &  1.49 &  1.45 &  vB99  &  \\  
   753 & K3\_V        &  0.94 &  0.07 &  0.01 &  0.01 &  6.81 &  5.83 &  5.01 &  4.49 &  4.07 &  3.45 & -9.99 & -9.99 & KTFS04, 19  &  \\  
   824 & K1.5\_III    &  1.88 &  0.11 &  0.05 &  0.03 &  5.63 &  4.52 &  3.72 &  3.14 &  2.70 &  2.02 &  1.98 &  1.92 & Nord99   &  \\  
   834 & K3\_Ib       &  5.38 &  0.05 &  0.92 &  0.35 &  5.48 &  3.79 &  2.56 &  1.67 &  1.05 &  0.09 & -0.11 & -0.11 & MAH03   &  \\  
   843 & K5\_III      &  4.06 &  0.04 &  0.12 &  0.08 &  6.09 &  4.53 &  3.32 &  2.37 &  1.76 &  0.78 &  0.64 &  0.60 & MAH03   &  \\  
   867 & M6\_III      & 10.30 &  0.21 &  0.10 &  0.01 &  7.44 &  5.93 &  3.51 &  1.34 &  0.24 & -1.05 & -1.39 & -1.49 & WF87  &  \\  
   882 & K2\_III      &  2.08 &  0.07 &  0.10 &  0.06 &  6.17 &  4.93 &  4.05 &  3.41 & -9.99 &  2.09 &  1.98 &  1.99 & vB99   &  \\  
   911 & M1.5\_III    & 13.23 &  0.16 &  0.06 &  0.05 &  4.17 &  2.53 &  1.18 &  0.02 & -0.56 & -1.64 & -1.89 & -1.92 & MJS91   &  \\  
   921 & M4\_II       & 16.56 &  0.17 &  0.09 &  0.06 &  5.04 &  3.39 &  1.59 & -0.03 & -0.78 & -1.93 & -2.19 & -2.30 & MAH03   &  \\  
   951 & K2\_III      &  1.87 &  0.13 &  0.01 &  0.01 &  5.40 &  4.37 &  3.60 &  3.09 &  2.69 &  2.08 &  1.97 &  1.94 & Nord99   &  \\  
  1017 & F5\_Iab      &  3.10 &  0.02 &  0.32 &  0.07 &  2.27 &  1.79 &  1.34 &  1.01 &  0.87 &  0.56 &  0.44 &  0.37 & Nord99   &  \\  
  1084 & K2\_V        &  2.15 &  0.03 &  0.00 &  0.01 &  4.61 &  3.73 &  3.01 &  2.54 &  2.24 &  1.70 &  1.59 &  1.39 & KTFS04   &  \\  
  1136 & K0\_IV       &  2.39 &  0.03 &  0.01 &  0.01 &  4.46 &  3.54 &  2.82 &  2.32 &  1.96 &  1.40 &  1.36 &  1.38 & KTFS04  &  \\  
  1231 & M1\_III      &  9.33 &  0.17 &  0.17 &  0.02 &  4.54 &  2.94 &  1.68 &  0.68 &  0.13 & -0.88 & -1.07 & -1.14 & MAH03   &  \\  
  1256 & K0\_III      &  1.69 &  0.08 &  0.08 &  0.03 &  5.44 &  4.37 &  3.58 &  3.05 &  2.63 &  1.97 & -9.99 & -9.99 & Nord99   &  \\  
  1318 & K3\_III      &  1.74 &  0.03 &  0.16 &  0.02 &  6.04 &  4.86 &  4.01 &  3.44 &  2.94 &  2.23 & 2.16 & 2.09 & KNB03   &  \\  
  1325 & K1\_V        &  1.65 &  0.06 &  0.02 &  0.01 &  5.25 &  4.43 &  3.74 &  3.29 &  2.95 &  2.40 & -9.99 & -9.99 & KTFS04  &  \\  
  1326 & M1.5\_V      &  1.00 &  0.05 &  0.01 &  0.01 &  9.63 &  8.07 &  6.72 &  5.53 &  4.85 &  4.01 & -9.99 & -9.99 & KTFS04, 19  &  \\  
  1343 & G8\_III      &  1.44 &  0.13 &  0.45 &  0.23 &  5.87 &  4.93 & -9.99 & -9.99 & -9.99 &  2.76 & -9.99 & -9.99 & vB99   &  \\  
  1373 & K0\_III      &  2.29 &  0.03 &  0.06 &  0.03 &  4.75 &  3.76 &  3.03 &  2.56 &  2.23 &  1.64 &  1.52 &  1.58 & Nord01    &  \\  
  1409 & G9.5\_III    &  2.67 &  0.03 &  0.06 &  0.03 &  4.55 &  3.54 &  2.81 &  2.31 &  1.94 &  1.33 &  1.26 &  1.19 & MAH03   &  \\  
  1457 & K5\_III      & 21.10 &  0.21 &  0.03 &  0.02 &  2.40 &  0.86 & -0.37 & -1.31 & -1.84 & -2.81 & -3.08 & -3.02 & MAH03  &  \\  
  1533 & K4\_III      &  2.93 &  0.08 &  0.33 &  0.12 &  6.31 &  4.88 &  3.80 &  2.95 & -9.99 &  1.46 &  1.29 &  1.23 & vB99, 8   &  \\  
  1551 & K3\_III      &  2.79 &  0.06 &  0.35 &  0.13 &  6.18 &  4.77 &  3.68 &  2.90 & -9.99 &  1.48 &  1.25 &  1.01 & vB99   &  \\  
  1577 & K3\_II       &  7.50 &  0.08 &  0.95 &  0.48 &  4.22 &  2.69 &  1.63 &  0.81 &  0.27 & -0.63 & -0.82 & -0.84 & MAH03   &  \\  
  1601 & K2\_II       &  2.78 &  0.06 &  0.51 &  0.21 &  5.89 &  4.49 &  3.44 &  2.74 & -9.99 &  1.41 &  1.26 &  1.22 & MAH03   &  \\  
  1605 & A8\_Iab      &  2.17 &  0.03 &  1.07 &  0.40 &  3.53 &  2.99 &  2.47 &  2.02 &  1.82 &  1.46 &  1.62 &  1.33 & Nord01   &  \\  
  1713 & B8\_Iab      &  2.55 &  0.05 &  0.20 &  0.09 &  0.10 &  0.13 &  0.13 &  0.14 &  0.22 &  0.20 & -0.01 & -0.13 & HBD74   &  \\  
  1739 & G8\_III      &  1.42 &  0.13 &  0.24 &  0.12 &  5.84 &  4.90 & -9.99 & -9.99 & -9.99 &  2.89 &  2.75 &  2.70 & vB99  &  \\  
  1791 & B7\_III      &  1.59 &  0.11 &  0.07 &  0.02 &  1.52 &  1.65 &  1.66 &  1.76 &  1.97 &  2.05 & -9.99 & -9.99 & vB99  &  \\  
  1845 & M2\_Iab      & 10.00 &  0.30 &  0.89 &  0.45 &  6.41 &  4.35 &  2.59 &  1.14 &  0.32 & -1.25 & -1.37 & -1.72 & WF87 & eliminated: $m_{12}-m_{25}$\\
  1865 & F0\_Ib       &  1.77 &  0.09 &  0.15 &  0.01 &  2.77 &  2.57 &  2.35 &  2.14 &  2.05 &  1.87 &  1.76 &  1.74 & Nord99   &  \\  
  1995 & G8\_III      &  1.97 &  0.08 &  0.07 &  0.02 &  5.47 &  4.53 &  3.80 &  3.31 &  2.92 &  2.34 & -9.99 & -9.99 & vB99   &  \\  
  2012 & G9.5\_III    &  2.79 &  0.06 &  0.07 &  0.02 &  5.11 &  3.97 &  3.15 &  2.59 &  2.18 &  1.51 &  1.36 &  1.39 & vB99   &  \\  
  2061 & M1\_Iab      & 45.20 &  0.20 &  0.10 &  0.01 &  2.29 &  0.40 & -1.19 & -2.37 & -2.93 & -4.01 & -5.14 & -5.66 & Dea96 & eliminated: $m_{12}-m_{25}$\\
\hline
\end{tabular}
\end{table*}

\setcounter{table}{ 0 }
\begin{table*}
\caption{Contiued}  
%\footnotesize
\setlength{\tabcolsep}{1.0mm}

\begin{tabular}{rlrcrcrrrrrrrrll} \hline

ID    & Type        & $\theta$ & $\sigma_{\theta}$ & $A_{\rm V}$ & $\sigma_{\rm A_V}$ & $B$ & $V$ & $R$ & $I$ & $J$ & $K$ & $[12]$ & $[25]$ & References & Remarks \\ \hline
  2091 & M3\_II       &  9.56 &  0.10 &  0.48 &  0.11 &  5.97 &  4.25 &  2.56 &  1.08 &  0.30 & -0.85 & -1.05 & -1.16 & MAH03   &  \\  
  2189 & M1\_III      &  2.94 &  0.06 &  0.30 &  0.09 &  7.44 &  5.78 & -9.99 & -9.99 & -9.99 &  1.47 &  1.30 &  1.23 & vB99   &  \\  
  2216 & M3\_III      & 11.79 &  0.12 &  0.18 &  0.05 &  4.89 &  3.28 &  1.79 &  0.48 & -0.23 & -1.31 & -1.67 & -1.74 & MAH03   &  \\  
  2219 & G8.5\_IIIb   &  2.16 &  0.09 &  0.05 &  0.03 &  5.36 &  4.35 &  3.55 &  3.01 &  2.59 &  1.94 &  1.72 &  1.82 & vB99   &  \\  
  2286 & M3\_III      & 15.12 &  0.15 &  0.12 &  0.04 &  4.51 &  2.87 &  1.30 & -0.08 & -0.73 & -1.85 & -2.18 & -2.22 & MAH03  &  \\  
  2421 & A0\_IV       &  1.38 &  0.09 &  0.03 &  0.02 &  1.92 &  1.92 &  1.86 &  1.87 &  1.90 &  1.90 & -9.99 & -9.99 & KTFS04  &  \\  
  2473 & G8\_Ib       &  4.73 &  0.03 &  0.23 &  0.07 &  4.38 &  2.98 &  2.02 &  1.41 &  0.99 &  0.22 &  0.01 &  0.03 & Nord99   &  \\  
  2491 & A1\_V        &  6.01 &  0.02 &  0.00 &  0.01 & -1.46 & -1.46 & -1.46 & -1.43 & -1.34 & -1.31 & -1.36 & -1.38 & KTFS04   &  \\  
  2630 & G5\_Ib-II    &  1.50 &  0.21 &  0.32 &  0.09 &  6.13 &  5.18 & -9.99 & -9.99 & -9.99 &  3.40 &  3.00 &  2.82 & vB99, 16 &  \\  
  2696 & K4\_Iab      &  2.92 &  0.10 &  0.07 &  0.03 &  6.37 &  4.93 &  3.84 &  3.07 & -9.99 &  1.51 &  1.41 &  1.36 & vB99, 8 &  \\  
  2805 & K0\_IIIa     &  1.94 &  0.17 &  0.10 &  0.04 &  6.46 &  5.22 &  4.40 &  3.80 &  3.10 &  2.49 &  2.38 &  2.36 & vB99, 8, 11 &  \\  
  2943 & F5\_IV-V     &  5.49 &  0.04 &  0.00 &  0.01 &  0.79 &  0.37 & -0.05 & -0.28 & -0.40 & -0.64 & -0.71 & -0.72 & KTFS04  &  \\  
  2973 & K1\_III\_SB  &  2.31 &  0.05 &  0.00 &  0.01 &  5.41 &  4.29 &  3.37 &  2.79 &  2.35 &  1.65 &  1.30 &  1.46 & Nord99   &  \\  
  2990 & K0\_IIIb     &  7.98 &  0.08 &  0.00 &  0.01 &  2.14 &  1.14 &  0.39 & -0.11 & -0.49 & -1.11 & -1.21 & -1.19 & MAH03   &  \\  
  3249 & K4\_III      &  5.03 &  0.04 &  0.00 &  0.01 &  5.01 &  3.53 &  2.41 &  1.63 &  1.07 &  0.14 &  0.03 & -0.08 & Nord99   &  \\  
  3547 & G9\_II-III   &  3.18 &  0.09 &  0.03 &  0.02 &  4.10 &  3.10 &  2.39 &  1.90 &  1.48 &  0.87 &  0.80 &  0.76 & Nord01   &  \\  
  3576 & M3\_III      &  5.64 &  0.14 &  0.00 &  0.01 &  6.29 &  4.76 &  3.29 &  2.03 &  1.33 &  0.29 &  0.07 &  0.01 & MAH03   &  \\  
  3705 & K7\_III      &  7.54 &  0.08 &  0.00 &  0.01 &  4.68 &  3.13 &  1.90 &  1.00 &  0.39 & -0.61 & -0.82 & -0.79 & MAH03   &  \\  
  3748 & K2\_III      &  9.73 &  0.10 &  0.07 &  0.01 &  3.42 &  1.97 &  0.93 &  0.16 & -0.33 & -1.19 & -1.46 & -1.37 & MAH03   &  \\  
  3873 & G1\_III      &  2.60 &  0.05 &  0.03 &  0.01 &  3.79 &  2.98 &  2.33 &  1.93 &  1.63 &  1.15 &  0.96 &  1.03 & Nord01   &  \\  
  3950 & M2\_III      &  4.62 &  0.06 &  0.02 &  0.01 &  6.30 &  4.70 &  3.27 &  2.19 &  1.54 &  0.50 &  0.33 &  0.26 & Nord99   &  \\  
  3980 & K4\_III      &  3.33 &  0.04 &  0.04 &  0.01 &  5.82 &  4.37 &  3.24 &  2.47 &  1.91 &  1.04 &  0.92 &  0.90 & Nord99   &  \\  
  4050 & K3\_IIa      &  5.18 &  0.05 &  0.14 &  0.07 &  4.89 &  3.36 &  2.26 &  1.50 & -9.99 &  0.03 & -0.17 & -0.18 & KNB03, 20 &  \\  
  4069 & M0\_III\_SB  &  8.54 &  0.09 &  0.09 &  0.09 &  4.64 &  3.05 &  1.77 &  0.81 &  0.09 & -0.88 & -0.98 & -1.07 & MAH03   &  \\  
  4247 & K0\_III      &  2.54 &  0.03 &  0.03 &  0.02 &  4.87 &  3.83 &  3.00 &  2.46 &  2.07 &  1.39 &  1.28 &  1.28 & Nord99   &  \\  
  4301 & K0\_Iab\_SB  &  7.11 &  0.10 &  0.00 &  0.01 &  2.86 &  1.79 &  0.98 &  0.40 &  0.05 & -0.65 & -0.80 & -0.81 & Nord01   &  \\  
  4335 & K1\_III      &  4.12 &  0.04 &  0.05 &  0.04 &  4.15 &  3.01 &  2.17 &  1.60 &  1.16 &  0.44 &  0.30 &  0.27 & MAH03  &  \\  
  4377 & K3\_III\_SB  &  4.76 &  0.05 &  0.05 &  0.03 &  4.89 &  3.49 &  2.43 &  1.73 &  1.18 &  0.31 &  0.15 &  0.15 & MAH03  &  \\  
  4432 & K3.5\_III    &  3.21 &  0.03 &  0.02 &  0.01 &  6.31 &  4.55 &  3.40 &  2.52 &  2.10 &  1.16 &  1.08 &  0.99 & Nord99, 4, 5 &  \\  
  4434 & M0\_III      &  6.43 &  0.07 &  0.00 &  0.01 &  5.47 &  3.85 &  2.54 &  1.55 &  0.87 & -0.14 & -0.31 & -0.42 & MAH03   &  \\  
  4517 & M1\_III      &  6.26 &  0.10 &  0.04 &  0.02 &  5.54 &  4.04 &  2.78 &  1.77 &  1.12 &  0.09 & -0.05 & -0.15 & Nord01   &  \\  
  4518 & K0.5\_IIIb   &  3.23 &  0.02 &  0.06 &  0.04 &  4.90 &  3.72 &  2.84 &  2.24 &  1.72 &  0.95 &  0.82 &  0.76 & Nord99   &  \\  
  4534 & A3\_V        &  1.45 &  0.03 &  0.01 &  0.01 &  2.22 &  2.14 &  2.08 &  2.06 &  2.03 &  1.99 &  1.92 &  1.63 & KTFS04  &  \\  
  4546 & K3\_III      &  2.94 &  0.03 &  0.18 &  0.02 &  5.76 &  4.46 &  3.52 &  2.85 &  2.36 &  1.56 &  1.44 &  1.41 & KNB03   &  \\  
  4853 & B0.5\_IV     &  0.72 &  0.02 &  0.16 &  0.06 &  1.02 &  1.25 &  1.38 &  1.64 &  1.80 &  1.99 & -9.99 & -9.99 & HBD74   &  \\  
  4910 & M3\_III      & 10.71 &  0.11 &  0.04 &  0.01 &  4.97 &  3.38 &  1.85 &  0.52 & -0.17 & -1.25 & -1.49 & -1.57 & MAH03   &  \\  
  4932 & G8\_III      &  3.17 &  0.03 &  0.02 &  0.01 &  3.78 &  2.84 &  2.20 &  1.75 &  1.36 &  0.80 &  0.70 &  0.61 & Nord99   &  \\  
  5200 & K5.5\_III    &  4.72 &  0.05 &  0.03 &  0.01 &  5.59 &  4.07 &  2.87 &  2.00 &  1.35 &  0.37 &  0.22 &  0.20 & Nord99   &  \\  
  5215 & M2\_III      &  2.50 &  0.05 &  0.10 &  0.05 &  7.49 &  5.87 & -9.99 & -9.99 &  2.72 &  1.67 &  1.66 &  1.52 & vB99, 9   &  \\  
  5235 & G0\_IB       &  2.27 &  0.03 &  0.01 &  0.01 &  3.26 &  2.68 &  2.24 &  1.95 &  1.70 &  1.37 &  1.23 &  1.25 & KTFS04   &  \\  
  5299 & M4.5\_III    &  7.85 &  0.11 &  0.10 &  0.05 &  6.86 &  5.28 &  3.43 &  1.77 &  0.86 & -0.29 & -0.57 & -0.73 & Wea01 & \\  
  5340 & K1.5\_III    & 21.00 &  0.20 &  0.01 &  0.01 &  1.18 & -0.05 & -1.02 & -1.67 & -2.08 & -2.95 & -3.22 & -3.09 & QMB96   &  \\  
  5429 & K3\_III      &  3.80 &  0.12 &  0.05 &  0.04 &  4.89 &  3.59 &  2.67 &  2.02 &  1.46 &  0.66 &  0.50 &  0.48 & vB99   &  \\  
  5459 & G2\_V        &  8.51 &  0.02 &  0.00 &  0.01 &  0.72 &  0.01 & -9.90 & -9.90 & -1.15 & -1.38 & -1.83 & -1.83 & KTFS04, 21  &  \\  
  5460 & K1\_V        &  6.01 &  0.03 &  0.00 &  0.01 &  2.21 &  1.33 & -9.90 & -9.90 & -0.01 & -0.60 & -9.99 & -9.99 & KTFS04, 21  &  \\  
  5510 & M1\_III      &  2.07 &  0.05 &  0.10 &  0.07 &  7.86 &  6.28 & -9.99 & -9.99 &  3.07 &  2.05 &  1.96 &  1.87 & vB99, 9 &  \\  
  5563 & K4\_III      & 10.30 &  0.10 &  0.00 &  0.01 &  3.55 &  2.08 &  0.97 &  0.21 & -0.45 & -1.39 & -1.48 & -1.51 & MAH03, 9   &  \\  
  5568 & K4\_V        &  1.23 &  0.03 &  0.02 &  0.01 &  6.82 &  5.71 &  4.72 &  4.18 &  3.84 &  3.06 &  2.49 &  2.61 & KTFS04  &  \\  
  5589 & M4.5\_III    & 10.59 &  0.17 &  0.06 &  0.01 &  6.18 &  4.59 &  2.73 &  1.02 &  0.22 & -0.95 & -1.21 & -1.35 & MAH03   &  \\  
  5602 & G8\_III      &  2.47 &  0.04 &  0.06 &  0.04 &  4.47 &  3.50 &  2.85 &  2.41 &  1.93 &  1.34 &  1.27 &  1.26 & Nord01   &  \\  
  5638 & K2\_III      &  1.44 &  0.06 &  0.05 &  0.01 &  6.92 &  5.68 &  4.71 &  4.10 & -9.99 &  2.67 &  2.58 &  2.58 & Nord01   &  \\  
  5681 & G8\_III      &  2.76 &  0.03 &  0.04 &  0.01 &  4.44 &  3.49 &  2.76 &  2.25 &  1.87 &  1.22 &  1.14 &  1.04 & MAH03   &  \\  
  5709 & K0\_III      &  1.18 &  0.07 &  0.05 &  0.01 &  6.53 &  5.51 & -9.99 & -9.99 & -9.99 &  3.13 &  3.02 &  2.96 & vB99   &  \\  
  5745 & M1\_III      &  2.42 &  0.05 &  0.05 &  0.01 &  7.64 &  6.02 & -9.99 & -9.99 &  2.84 &  1.76 &  1.72 &  1.59 & vB99, 9 &  \\  
  5854 & K2\_III      &  4.85 &  0.05 &  0.06 &  0.05 &  3.81 &  2.64 &  1.83 &  1.27 &  0.76 &  0.07 &  0.00 & -0.04 & MAH03   &  \\  
  6056 & M0.5\_III    & 10.47 &  0.12 &  0.14 &  0.11 &  4.34 &  2.75 &  1.46 &  0.43 & -0.20 & -1.22 & -1.41 & -1.49 & MAH03   &  \\  
  6107 & M2\_III      &  3.68 &  0.10 &  0.10 &  0.03 &  6.80 &  5.20 & -9.99 & -9.99 &  1.92 &  0.84 &  0.72 &  0.66 & vB99, 9 &  \\  
  6132 & G8\_III\_SB  &  3.68 &  0.05 &  0.03 &  0.01 &  3.65 &  2.74 &  2.13 &  1.67 &  1.17 &  0.59 &  0.54 &  0.49 & Nord01   &  \\  
  6134 & M1.5\_Ib     & 39.76 &  0.40 &  0.73 &  0.27 &  2.75 &  0.91 & -0.64 & -1.87 & -2.74 & -3.79 & -4.73 & -4.65 & MAH03, 4, 21 &  \\  
  6146 & M6\_III      & 19.09 &  0.19 &  0.10 &  0.03 &  6.58 &  5.05 &  2.51 &  0.28 & -0.70 & -1.97 & -2.57 & -2.99 & MAH03  & eliminated: $m_{12}-m_{25}$ \\
  6148 & G7\_IIIa\_SB &  3.46 &  0.04 &  0.05 &  0.03 &  3.71 &  2.77 &  2.13 &  1.66 &  1.21 &  0.60 &  0.56 &  0.54 & MAH03   &  \\  
  6208 & K2\_III      &  1.24 &  0.06 &  0.07 &  0.04 &  7.38 &  6.06 & -9.99 & -9.99 & -9.99 &  2.94 &  2.83 &  2.83 & vB99  &  L.C. based on $M_{\rm V}$\\ 
  6212 & G0\_IV\_SB   &  2.27 &  0.03 &  0.02 &  0.01 &  3.46 &  2.81 &  2.30 &  1.98 &  1.70 &  1.30 &  1.22 &  1.19 & KTFS04   &  \\  
\hline
\end{tabular}
\end{table*}

\setcounter{table}{ 0 }
\begin{table*}  
\caption{Contiued}  
%\footnotesize
\setlength{\tabcolsep}{1.0mm}

\begin{tabular}{rlrcrcrrrrrrrrll} \hline

ID    & Type        & $\theta$ & $\sigma_{\theta}$ & $A_{\rm V}$ & $\sigma_{\rm A_V}$ & $B$ & $V$ & $R$ & $I$ & $J$ & $K$ & $[12]$ & $[25]$ & References & Remarks \\ \hline
  6220 & G7.5\_III    &  2.62 &  0.03 &  0.05 &  0.02 &  4.42 &  3.50 &  2.83 &  2.35 &  1.98 &  1.35 &  1.27 &  1.24 & MAH03   &  \\  
  6258 & M1\_III      &  2.32 &  0.06 &  0.10 &  0.03 &  7.31 &  5.72 & -9.99 & -9.99 &  2.82 &  1.78 &  1.59 &  1.64 & vB99, 9 &  \\  
  6406 & M5\_Ib-II    & 36.03 &  0.44 &  0.69 &  0.20 &  4.92 &  3.48 & -9.99 & -9.99 & -2.30 & -3.47 & -3.92 & -4.14 & MAH03   &  \\  
  6418 & K3\_Iab      &  5.20 &  0.03 &  0.10 &  0.03 &  4.60 &  3.16 &  2.20 &  1.48 &  0.85 & -0.02 & -0.18 & -0.20 & Nord99   &  \\  
  6536 & G2\_Iab      &  3.22 &  0.05 &  0.10 &  0.03 &  3.78 &  2.78 &  2.10 &  1.62 &  1.29 &  0.77 &  0.53 &  0.59 & MAH03   &  \\  
  6584 & M2\_III      &  2.26 &  0.05 &  0.17 &  0.01 &  7.61 &  6.03 & -9.99 & -9.99 &  2.89 &  1.81 &  1.68 &  1.67 & vB99, 9   &  \\  
  6623 & G5\_IV       &  1.93 &  0.03 &  0.01 &  0.01 &  4.17 &  3.42 &  2.89 &  2.51 &  2.18 &  1.77 & -9.99 & -9.99 & KTFS04   &  \\  
  6695 & K1\_IIaCN    &  3.17 &  0.03 &  0.11 &  0.01 &  5.22 &  3.87 &  2.97 &  2.34 &  1.83 &  1.03 &  0.87 &  0.89 & MAH03   &  \\  
  6705 & K5\_III      &  9.86 &  0.13 &  0.03 &  0.01 &  3.74 &  2.22 &  1.08 &  0.23 & -0.39 & -1.24 & -1.44 & -1.50 & MAH03   &  \\  
  6820 & K4\_III      &  1.54 &  0.06 &  0.32 &  0.02 &  7.59 &  6.12 & -9.99 & -9.99 & -9.99 &  2.63 &  2.49 &  2.41 & vB99   &  \\  
  7001 & A0\_V        &  3.28 &  0.01 &  0.01 &  0.01 &  0.03 &  0.03 &  0.07 &  0.10 &  0.02 &  0.02 & -0.01 & -0.16 & Cea01   &  \\  
  7139 & M4\_II       & 11.53 &  0.16 &  0.11 &  0.01 &  5.97 &  4.30 &  2.52 &  0.89 & -0.02 & -1.23 & -1.45 & -1.68 & MAH03   &  \\  
  7157 & M5\_III      & 18.02 &  0.22 &  0.07 &  0.01 &  5.66 &  4.08 &  2.04 &  0.13 & -0.85 & -2.07 & -2.39 & -2.54 & MAH03   &  \\  
  7176 & K1\_III      &  1.99 &  0.11 &  0.10 &  0.03 &  5.10 &  4.02 &  3.26 &  2.74 &  2.33 &  1.69 & -9.99 & -9.99 & Nord99   &  \\  
  7193 & K1\_III      &  2.42 &  0.01 &  0.19 &  0.07 &  5.11 &  4.02 &  3.23 &  2.69 &  2.23 &  1.44 &  1.43 &  1.42 & Aea01   &  \\  
  7237 & M0\_III      &  2.27 &  0.05 &  0.16 &  0.01 &  7.14 &  5.59 & -9.99 & -9.99 &  2.79 &  1.78 &  1.79 &  1.67 & vB99, 9   &  \\  
  7238 & M2\_III      &  2.40 &  0.05 &  0.19 &  0.01 &  7.61 &  6.06 & -9.99 & -9.99 &  2.80 &  1.70 &  1.64 &  1.50 & vB99, 9   &  \\  
  7310 & G9\_III      &  3.25 &  0.05 &  0.01 &  0.01 &  4.07 &  3.07 &  2.37 &  1.86 &  1.43 &  0.80 &  0.72 &  0.66 & MAH03   &  \\  
  7314 & K0\_III      &  2.23 &  0.09 &  0.19 &  0.01 &  5.62 &  4.37 &  3.50 &  2.91 &  2.37 &  1.66 & -9.99 & -9.99 & Nord99   &  \\  
  7328 & G9\_III      &  2.07 &  0.09 &  0.01 &  0.01 &  4.73 &  3.76 &  3.12 &  2.66 &  2.25 &  1.67 &  1.55 &  1.52 & Nord99   &  \\  
  7405 & M0\_III      &  4.46 &  0.05 &  0.12 &  0.04 &  5.95 &  4.45 &  3.24 &  2.27 &  1.55 &  0.55 &  0.38 &  0.33 & MAH03   &  \\  
  7417 & K3\_II       &  4.83 &  0.05 &  0.20 &  0.08 &  4.21 &  3.08 &  2.21 &  1.55 &  1.01 &  0.16 &  0.03 & -0.06 & MAH03   &  \\  
  7517 & G7\_III      &  1.18 &  0.06 &  0.04 &  0.04 &  5.86 &  4.90 & -9.99 & -9.99 & -9.99 &  2.73 &  2.66 &  2.78 & vB99   &  \\  
  7525 & K3\_II       &  7.08 &  0.05 &  0.23 &  0.06 &  4.24 &  2.72 &  1.65 &  0.90 &  0.30 & -0.59 & -0.68 & -0.81 & Nord99   &  \\  
  7536 & M2\_II       &  9.15 &  0.12 &  0.18 &  0.06 &  5.24 &  3.83 &  2.39 &  1.10 &  0.37 & -0.79 & -0.99 & -1.08 & MAH03   &  \\  
  7557 & A7\_V        &  3.46 &  0.04 &  0.01 &  0.01 &  0.98 &  0.76 &  0.62 &  0.48 &  0.39 &  0.26 &  0.24 &  0.18 & MAH03   &  \\  
  7570 & F6\_Iab\_SB  &  1.73 &  0.05 &  0.62 &  0.15 &  4.73 &  3.86 &  3.27 &  2.83 &  2.45 &  1.98 &  1.89 &  1.87 & Nord99, 1, 6 &  \\  
  7602 & G8\_IV       &  2.18 &  0.09 &  0.02 &  0.01 &  4.58 &  3.72 &  3.06 &  2.57 &  2.26 &  1.71 &  1.65 &  1.60 & Nord99  &  \\  
  7635 & M0\_III      &  6.22 &  0.06 &  0.06 &  0.01 &  5.04 &  3.47 &  2.27 &  1.35 &  0.79 & -0.16 & -0.39 & -0.42 & MAH03   &  \\  
  7735 & K2\_II\_SB   &  4.36 &  0.04 &  1.04 &  0.36 &  5.08 &  3.80 &  2.83 &  2.07 &  1.41 &  0.49 &  0.38 &  0.18 & MAH03   &  \\  
  7751 & K3\_Ib\_SB   &  5.42 &  0.05 &  0.26 &  0.09 &  5.50 &  3.98 &  2.78 &  1.86 &  1.19 &  0.16 & -0.04 & -0.09 & MAH03   &  \\  
  7759 & K5\_Iab      &  1.19 &  0.06 &  0.64 &  0.23 &  6.90 &  5.25 &  4.07 &  3.20 & -9.99 &  1.45 &  1.39 &  1.36 & vB99, 2  &  \\  
  7796 & F8\_Iab      &  3.02 &  0.03 &  0.94 &  0.34 &  2.90 &  2.23 &  1.74 &  1.40 &  1.16 &  0.72 &  0.67 & -9.99 & MAH03  &  \\  
  7806 & K3\_III      &  2.99 &  0.08 &  0.17 &  0.06 &  5.77 &  4.44 &  3.43 &  2.76 &  2.29 &  1.44 &  1.26 &  1.26 & vB99  &  \\  
  7834 & F5\_Iab      &  1.43 &  0.08 &  0.26 &  0.12 &  4.42 &  4.02 &  3.66 &  3.43 &  3.21 &  2.94 &  2.83 &  2.85 & vB99, 13   &  \\  
  7847 & F5\_Iab      &  1.23 &  0.14 &  0.97 &  0.35 &  7.19 &  6.18 &  5.33 &  4.64 &  4.17 &  3.54 &  3.26 &  2.18 & vB99   &  \\  
  7924 & A2\_Ia       &  2.42 &  0.06 &  1.39 &  0.63 &  1.34 &  1.25 &  1.14 &  1.04 &  1.00 &  0.89 & -9.99 & -9.99 & MAH03   &  \\  
  7942 & G9.5\_III    &  2.35 &  0.05 &  0.08 &  0.04 &  5.29 &  4.23 &  3.46 &  2.93 &  2.56 &  1.87 &  1.68 &  1.65 & vB99   &  \\  
  7949 & K0\_III      &  4.62 &  0.04 &  0.03 &  0.02 &  3.49 &  2.46 &  1.73 &  1.19 &  0.77 &  0.11 & -0.04 & -0.01 & MJS91   &  \\  
  7957 & K0\_IV       &  2.67 &  0.04 &  0.06 &  0.03 &  4.35 &  3.43 &  2.76 &  2.27 &  1.90 &  1.28 &  1.12 &  1.16 & KTFS04   &  \\  
  7995 & G7\_III      &  1.54 &  0.29 &  0.09 &  0.04 &  5.43 &  4.61 &  3.93 &  3.47 &  3.11 &  2.63 &  2.57 &  2.57 & vB99, 17  & \\  
  8008 & K4\_III      &  2.46 &  0.05 &  0.25 &  0.12 &  6.50 &  5.02 & -9.99 & -9.99 &  2.47 &  1.58 &  1.54 &  1.47 & vB99, 9 &  \\  
  8044 & M3\_III      &  3.33 &  0.08 &  0.11 &  0.03 &  7.26 &  5.65 & -9.99 & -9.99 & -9.99 &  1.21 &  1.11 &  0.96 & vB99   &  \\  
  8079 & K4.5\_Ib\_SB &  5.56 &  0.04 &  0.53 &  0.24 &  5.35 &  3.70 &  2.50 &  1.60 &  1.00 & -0.05 & -9.99 & -9.99 & Nord99   &  \\  
  8115 & G8\_III      &  2.82 &  0.03 &  0.06 &  0.03 &  4.19 &  3.20 &  2.50 &  2.02 &  1.65 &  1.09 &  1.03 &  0.94 & MAH03   &  \\  
  8225 & M1\_III      &  4.52 &  0.05 &  0.08 &  0.02 &  6.19 &  4.57 &  3.32 &  2.23 &  1.54 &  0.51 &  0.37 &  0.33 & MAH03, 14   &  \\  
  8252 & G8\_III      &  1.82 &  0.10 &  0.09 &  0.03 &  4.91 &  4.02 &  3.31 &  2.81 &  2.55 &  1.97 & -9.99 & -9.99 & Nord99  &  \\  
  8306 & M2\_III      &  3.24 &  0.07 &  0.07 &  0.05 &  7.09 &  5.49 & -9.99 & -9.99 & -9.99 &  1.31 &  1.11 &  1.09 & vB99  & \\  
  8308 & K2\_Ib       &  7.54 &  0.14 &  0.13 &  0.10 &  3.91 &  2.39 &  1.34 &  0.58 &  0.03 & -0.81 & -1.01 & -1.05 & Nord01   &  \\  
  8316 & M2\_Iae      & 20.58 &  0.48 &  2.00 &  0.70 &  6.43 &  4.17 &  2.07 &  0.31 & -0.52 & -1.65 & -3.75 & -4.52 & MAH03  & eliminated: $m_{12}-m_{25}$ \\
  8387 & K4.5\_V      &  1.89 &  0.02 &  0.01 &  0.01 &  5.74 &  4.69 &  3.80 &  3.25 &  2.83 &  2.18 & -9.99 & -9.99 & KTFS04   &  \\  
  8414 & G2\_Ib       &  3.08 &  0.03 &  0.15 &  0.03 &  3.90 &  2.93 &  2.27 &  1.80 &  1.48 &  0.96 &  0.82 &  0.84 & Nord99   &  \\  
  8465 & K1.5Ib\_SB   &  5.23 &  0.05 &  0.69 &  0.24 &  4.90 &  3.35 &  2.27 &  1.49 &  0.97 &  0.11 &  0.07 & -0.14 & MAH03   &  \\  
  8538 & G8.5\_III    &  1.92 &  0.02 &  0.17 &  0.06 &  5.46 &  4.44 &  3.69 &  3.12 &  2.69 &  2.07 &  1.88 &  1.86 & Nord99   &  \\  
  8555 & K2\_II       &  1.52 &  0.06 &  0.15 &  0.08 &  7.43 &  5.98 & -9.99 & -9.99 & -9.99 &  2.62 &  2.47 &  2.38 & vB99  &  L.C. based on $M_{\rm V}$ \\
  8571 & F5\_Iab      &  1.52 &  0.02 &  0.91 &  0.32 &  4.35 &  3.70 & -9.99 &  2.95 &  2.69 &  2.31 &  2.23 &  2.02 & Nord99, 17  &  \\  
  8621 & M4\_III      &  7.39 &  0.12 &  0.42 &  0.15 &  6.67 &  5.09 & -9.99 & -9.99 & -9.99 & -0.22 & -0.39 & -0.52 & Wea01   &  \\  
  8632 & K2\_III      &  2.63 &  0.05 &  0.15 &  0.05 &  5.79 &  4.46 &  3.54 &  2.86 &  2.36 &  1.51 &  1.40 &  1.26 & Nord99  &  \\  
  8650 & G2\_III\_SB  &  3.26 &  0.07 &  0.04 &  0.02 &  3.81 &  2.95 &  2.31 &  1.83 &  1.46 &  0.93 &  0.78 &  0.80 & Nord01   &  \\  
  8667 & G8\_III      &  2.39 &  0.05 &  0.09 &  0.07 &  5.02 &  3.94 &  3.18 &  2.67 &  2.25 &  1.65 &  1.53 &  1.50 & MAH03  &  \\  
  8684 & G8\_III      &  2.50 &  0.04 &  0.06 &  0.01 &  4.42 &  3.48 &  2.80 &  2.33 &  2.01 &  1.43 &  1.32 &  1.31 & MAH03   &  \\  
  8698 & M2\_III      &  8.19 &  0.10 &  0.07 &  0.02 &  5.44 &  3.79 &  2.37 &  1.18 &  0.43 & -0.70 & -9.99 & -9.99 & MAH03   &  \\  
  8728 & A3\_V        &  2.23 &  0.02 &  0.02 &  0.01 &  1.25 &  1.16 &  1.10 &  1.08 &  1.02 &  0.97 & -9.99 & -9.99 & KTFS04  &  \\  
\hline
\end{tabular}
\end{table*}

\setcounter{table}{ 0 }
\begin{table*}  
\caption{Contiued}  
%\footnotesize
\setlength{\tabcolsep}{1.0mm}

\begin{tabular}{rlrcrcrrrrrrrrll} \hline

ID    & Type        & $\theta$ & $\sigma_{\theta}$ & $A_{\rm V}$ & $\sigma_{\rm A_V}$ & $B$ & $V$ & $R$ & $I$ & $J$ & $K$ & $[12]$ & $[25]$ & References & Remarks \\ \hline
  8775 & M3\_II-III   & 17.98 &  0.18 &  0.03 &  0.02 &  4.15 &  2.50 &  0.93 & -0.38 & -1.04 & -2.20 & -2.44 & -2.53 & MJS91   &  \\  
  8796 & K0\_Iab\_SB  &  2.34 &  0.05 &  0.15 &  0.03 &  6.12 &  4.77 &  3.80 &  3.12 & -9.99 &  1.72 &  1.62 &  1.54 & MAH03  &  \\  
  8923 & G7\_III      &  1.61 &  0.17 &  0.07 &  0.02 &  5.50 &  4.56 &  3.82 &  3.37 &  3.00 &  2.44 &  2.35 &  2.25 & Nord99   &  \\  
  8930 & G8\_III      &  1.79 &  0.07 &  0.13 &  0.08 &  6.24 &  5.22 & -9.99 & -9.99 & -9.99 &  2.73 &  2.60 &  2.69 & vB99  &  \\  
  8942 & M3\_III      &  3.70 &  0.10 &  0.15 &  0.03 &  7.79 &  6.06 & -9.99 & -9.99 & -9.99 &  0.99 &  0.80 &  0.65 & vB99  &  \\  
  8953 & M1\_III      &  2.53 &  0.05 &  0.15 &  0.03 &  8.18 &  6.45 & -9.99 & -9.99 & -9.99 &  1.86 &  1.74 &  1.66 & vB99  &  \\  
  8961 & G8\_III      &  2.66 &  0.08 &  0.05 &  0.02 &  4.84 &  3.82 &  3.04 &  2.47 &  1.97 &  1.32 &  1.19 &  1.17 & Nord99   &  \\  
  8974 & K1\_V        &  3.24 &  0.03 &  0.01 &  0.01 &  4.24 &  3.21 &  2.46 &  1.95 & -9.99 &  0.89 &  0.74 &  0.79 & Nord99   &  \\  
  9035 & M2\_III      &  2.46 &  0.05 &  0.15 &  0.03 &  7.71 &  6.11 & -9.99 & -9.99 & -9.99 &  2.01 &  1.78 &  1.67 & vB99   &  \\  
  9045 & G2\_Ia       &  2.47 &  0.05 &  2.47 &  0.89 &  5.85 &  4.49 &  3.63 &  2.89 &  2.59 &  2.03 &  0.77 &  0.53 & Nord99   &  \\  
  9055 & M2\_III      &  2.41 &  0.05 &  0.15 &  0.03 &  7.75 &  6.15 & -9.99 & -9.99 & -9.99 &  1.96 &  1.78 & -9.99 & vB99, 12 & \\  
   551 & M5.5\_V      &  1.05 &  0.08 &  0.00 &  0.01 & 13.02 & 11.05 &  8.68 &  6.42 &  5.33 &  4.37 &  3.89 &  4.25 & KTFS04  & GJ 551 \\  
   555 & M4\_III      &  8.66 &  0.20 &  0.10 &  0.03 &  6.00 &  4.41 &  2.68 &  1.17 &  0.52 & -0.64 & -0.84 & -0.98 & WAK04, 21 & GJ 555 \\  
   887 & M0.5\_V      &  1.39 &  0.04 &  0.01 &  0.01 &  8.82 &  7.36 & -9.90 & -9.90 &  4.20 &  3.36 &  3.02 &  2.94 & KTFS04  & GJ 887 \\  
 33793 & M1\_V        &  0.69 &  0.06 &  0.01 &  0.01 & 10.40 &  8.55 & -9.90 & -9.90 &  5.77 &  5.05 &  4.66 & -9.99 & KTFS04  & HD 33793 \\  
 36395 & M1.5\_V      &  1.15 &  0.11 &  0.01 &  0.01 &  9.44 &  7.97 &  6.53 &  5.39 &  4.72 &  3.82 &  3.60 &  3.79 & KTFS04  & HD 36395 \\  
 87937 & M4\_V        &  1.00 &  0.04 &  0.00 &  0.01 & 11.28 &  9.54 &  7.71 &  6.03 &  5.45 &  4.53 &  3.90 & -9.99 & KTFS04  & HIP 87937 \\  
 88230 & K7\_V        &  1.29 &  0.04 &  0.00 &  0.01 &  7.94 &  6.59 &  5.36 &  4.56 &  3.97 &  3.15 &  3.11 &  2.98 & KTFS04  & HD 88230 \\  
 95735 & M2\_V        &  1.44 &  0.03 &  0.01 &  0.01 &  9.00 &  7.49 &  5.99 &  4.80 &  4.10 &  3.31 &  3.04 &  2.86 & KTFS04  & HD 95735 \\  
+20092 & M2\_III      &  2.41 &  0.07 &  2.31 &  1.16 & 10.26 &  8.47 & -9.99 & -9.99 &  3.28 &  2.11 &  1.81 &  1.78 & vB99, 18 &  \\  
+20282 & M5\_II-III   & 10.00 &  0.30 &  0.05 &  0.01 &  7.97 &  6.65 & -9.99 & -9.99 &  0.20 & -1.04 & -1.66 & -2.46 & Dea96, 15, 18 & eliminated: $m_{12}-m_{25}$\\
+30086 & M3\_III      &  2.41 &  0.05 &  2.31 &  1.16 &  9.51 &  7.75 & -9.99 & -9.99 &  3.19 &  2.09 &  1.83 &  1.69 & vB99, 12, 18 &  \\  
+30105 & M2\_III      &  2.87 &  0.06 &  1.78 &  0.89 &  8.16 &  6.38 & -9.99 & -9.99 & -9.99 &  1.59 &  1.39 &  1.33 & vB99   &  \\  
+30115 & M1.5\_Ia     &  3.11 &  0.07 &  1.19 &  0.34 &  9.55 &  7.35 & -9.99 & -9.99 & -9.99 &  1.74 &  1.23 &  0.78 & vB99 & eliminated: $m_{12}-m_{25}$\\  
+30257 & M7.5\_III    & 18.80 &  0.50 &  0.10 &  0.07 &  9.24 &  7.97 & -9.99 & -9.99 & -9.99 & -1.85 & -3.29 & -4.11 & Dea96, 18 & eliminated: $m_{12}-m_{25}$\\
+30309 & M1\_II       &  2.95 &  0.06 &  0.26 &  0.02 &  8.01 &  6.42 &  4.84 &  3.42 & -9.99 &  1.41 &  1.19 &  1.15 & vB99   &  \\  
+30412 & M3\_Ib       &  2.91 &  0.06 &  1.50 &  0.55 & 10.21 &  7.99 &  5.83 &  4.06 &  3.06 &  1.67 &  1.25 &  0.72 & vB99 & eliminated: $m_{12}-m_{25}$\\  
+30438 & M5\_II       &  2.54 &  0.06 &  0.84 &  0.38 & 11.07 &  9.30 & -9.99 & -9.99 & -9.99 &  2.33 &  1.93 &  1.70 & vB99   &  \\  
+30465 & M3\_III      &  2.21 &  0.05 &  0.31 &  0.14 &  8.53 &  6.89 & -9.99 & -9.99 & -9.99 &  2.09 &  1.81 &  1.78 & vB99, 18 &  \\  
+30468 & M2\_III      &  2.14 &  0.05 &  0.29 &  0.13 &  8.91 &  7.10 & -9.99 & -9.99 &  3.37 &  2.15 &  1.91 &  1.82 & vB99, 10, 18 &  \\  
+40018 & M3\_III      &  2.39 &  0.09 &  0.24 &  0.08 &  8.77 &  7.08 & -9.99 & -9.99 & -9.99 &  2.18 &  1.96 &  1.80 & vB99  & \\  
+40022 & M4\_III      &  2.49 &  0.05 &  0.24 &  0.08 &  8.61 &  7.00 & -9.99 & -9.99 & -9.99 &  1.87 &  1.76 &  1.68 & vB99  & \\  
+40254 & K5\_III      &  1.57 &  0.07 &  0.10 &  0.05 &  8.50 &  6.90 & -9.99 & -9.99 & -9.99 &  2.85 &  2.76 &  2.62 & vB99  & \\  
+40337 & K5\_II       &  1.34 &  0.06 &  0.20 &  0.19 &  9.40 &  7.76 & -9.99 & -9.99 & -9.99 &  2.86 &  2.88 &  2.59 & vB99  & L.C. based on $M_{\rm V}$\\  
Sun    & G2\_V    & 1919260. & (1919.) & 0.00 &  0.00 & -26.12 & -26.75 & -27.12 & -27.48 & -27.86 & -28.22 & -9.99 & -9.99 & 22 & \\
\hline
\end{tabular}

\noindent
References:\\

HBD74  = Hanbury Brown et al. (1974)
WF87   = White \& Feierman (1987)
%HJM89  = Hutter et al. (1989)
MJS91  = Mozurkewich et al. (1991)
%DB93   = Di Benedetto (1993)
Dea96  = Dyck et al. (1996)
QMB96  = Quirrenbach et al. (1996)
Nord99 = Nordgren et al. (1999)
vB99   = van Belle et al. (1999)
WDH00  = Weiner et al. (2000)
Aea01  = Armstrong et al. (2001)
Cea01  = Ciardi et al. (2001)
Nord01 = Nordgren et al. (2001)
Wea01  = Wittkowski et al. (2001)
MAH03  = Mozurkewich et al. (2003)
KNB03  = Kervella et al. (2003)
KTFS04 = Kervella et al. (2004)
WAK04  = Wittkowski et al. (2004)

 1 = Wisniewski \& Johnson (1968)
 2 = Fernie (1972)
 3 = Gehrz \& Hackwell (1974)
 4 = Honeycutt et al. (1977)
 5 = Barnes et al. (1978)
 6 = Moffett \& Barnes (1980)
 7 = Scharlach \& Craine (1980)
 8 = Fernie (1983)
 9 = McWilliam \& Lambert (1984)
10 = Arevalo et al. (1988)
11 = Smith (1988)
12 = Noguchi (1989) 
13 = Blackwell et al. (1990)
14 = Fluks et al. (1994)
15 = Kerschbaum (1995)
16 = Richichi et al. (1996)
17 = Barnes et al. (1997)
18 = Richichi \& Percheron (2002)
19 = Morel \& Magnenat (1978)
20 = Cohen et al. (1999)
21 = Ducati (2002)
22 = Colina et al. (1996)

\label{TAB-LIST}
\end{table*}

\begin{figure}
\includegraphics[width=85mm]{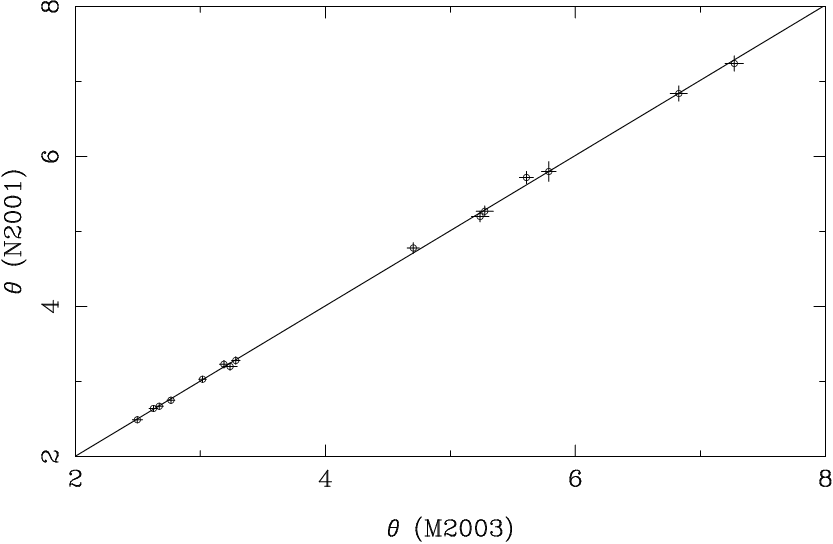}
\caption[]{
Limb-darkened angular diameters (in mas) from Nordgren et al. (2001)
plotted against those from Mozurkewich et al. (2003) for the stars in
common.  The line is a least-squares fit to the 15 data points
plotted: $\theta_{\rm N2001} = (1.002 \pm 0.007) \; \theta_{\rm M2003}
+ (0.000 \pm 0.031)$ (rms = 0.039).
}
\label{II}
\end{figure}

\begin{figure}
\includegraphics[width=85mm]{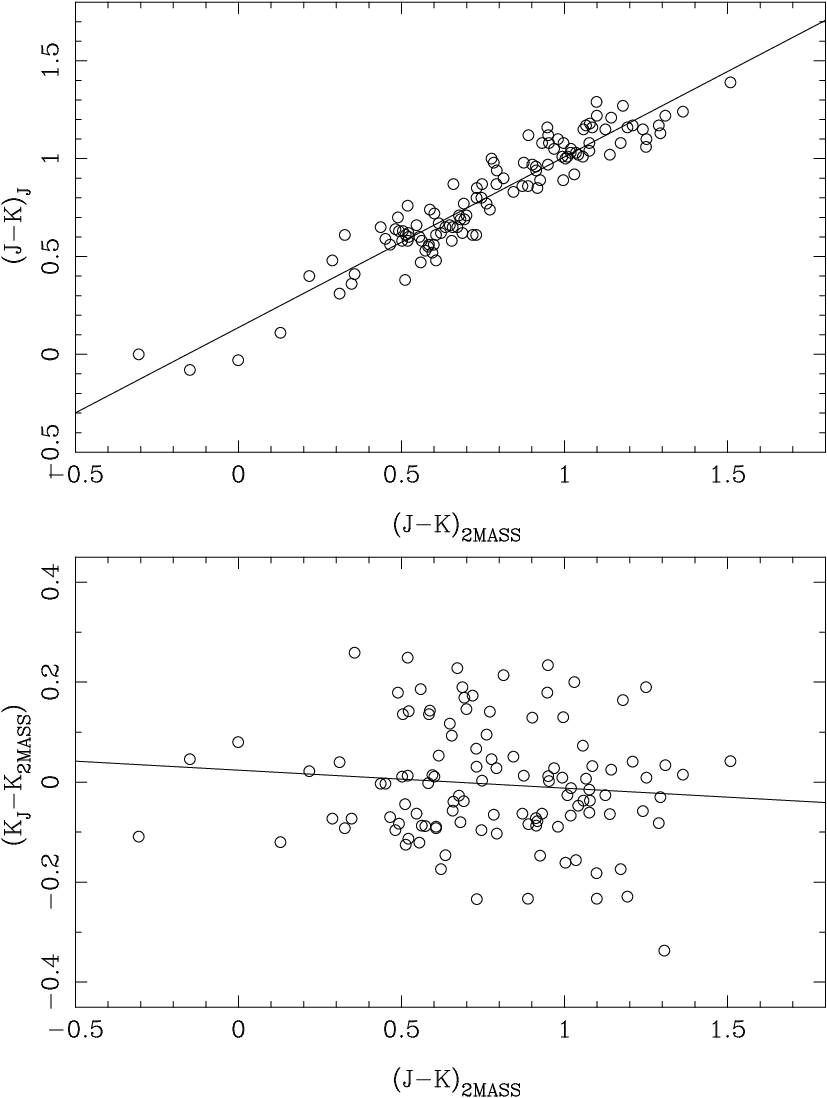}
\caption[]{
Top: ($J-K$) versus $(J-K)$ of \M. The line is a least-squares fit to
the 115 data points plotted: ($J-K) = (0.872 \pm 0.029) \; (J-K)_{\rm 2mass} 
+ (0.137 \pm 0.026)$ (rms = 0.096). Bottom: ($K-K_{\rm 2mass}$) versus
$(J-K)$ of \M.  The line is a least-squares fit: ($K-K_{\rm 2mass}) =
(-0.036 \pm 0.035) \; (J-K)_{\rm 2mass} + (0.024 \pm 0.029)$ (rms = 0.12).
}
\label{JKJK}
\end{figure}

\begin{figure}
\includegraphics[width=85mm]{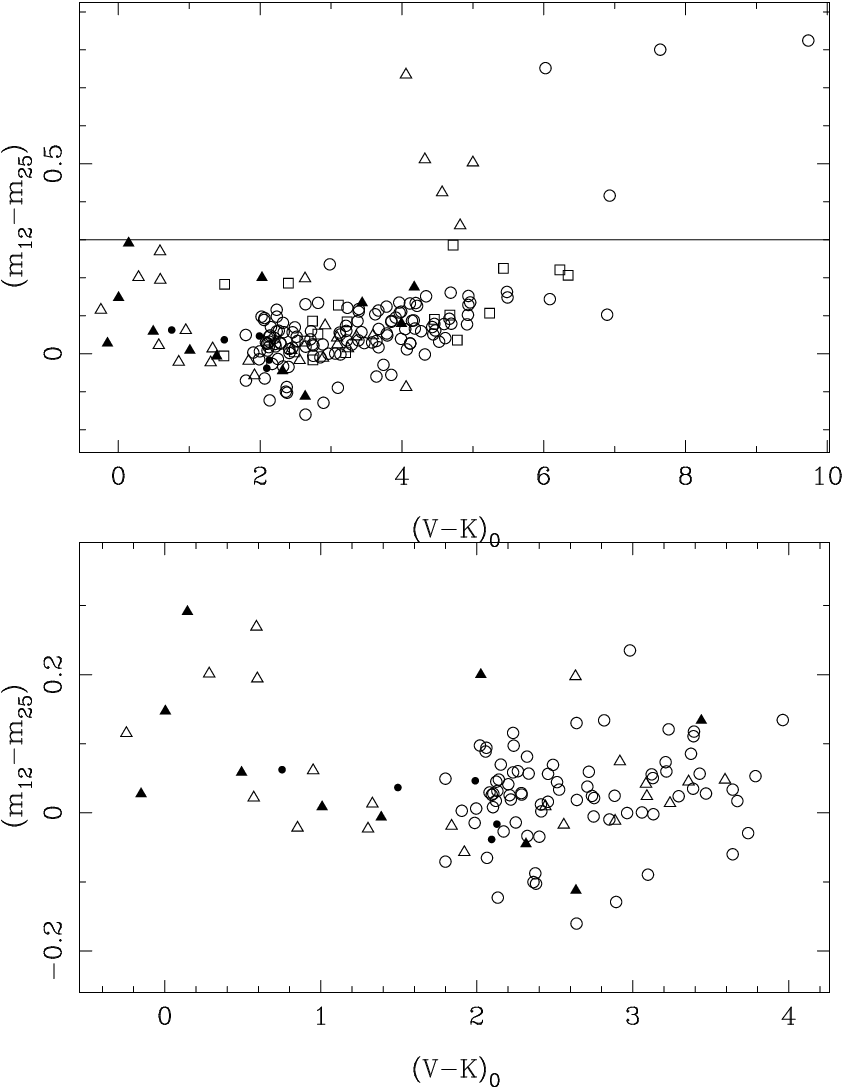}
\caption[]{
Top: ($m_{12}-m_{25}$) versus $(V-K)$ colour-colour relation. Symbols
are giants ($\circ$, defined as luminosity class {\sc iii} objects),
luminous giants ($\Box$, defined as luminosity class {\sc ii}
objects), supergiants ($\triangle$, defined as luminosity class {\sc
i} objects), sub-giants ($\bullet$, defined as luminosity class {\sc
iv} objects) and dwarfs (filled triangle, defined as luminosity class
{\sc v} objects).  The 5 supergiants and 4 giants with
($m_{12}-m_{25}) > 0.30$ are all of spectral type M. The line
indicates the cut-off at ($m_{12}-m_{25}) = 0.30$. Below: as top panel,
but without any star of spectral type M.
}
\label{Fig-CC}
\end{figure}

\setcounter{table}{ 1 }
\begin{table}
\setlength{\tabcolsep}{0.6mm}
\caption{Objects with three or more independent angular diamter estimates.}
\begin{tabular}{rcccc} \hline

ID   & $\theta$ $\pm$ $\sigma_{\theta}$ &  $\theta$ $\pm$ $\sigma_{\theta}$ &  $\theta$ $\pm$ $\sigma_{\theta}$ &  $\theta$ $\pm$ $\sigma_{\theta}$  \\ 
     &         (mas)                 &     (mas)                 &     (mas)                 &     (mas)    \\ \hline
 168 &  5.608 $\pm$ 0.056 (1) & 5.72 $\pm$ 0.08 (2) & 5.60 $\pm$ 0.06 (3) &                    \\
 617 &  6.827 $\pm$ 0.068 (1) & 6.84 $\pm$ 0.10 (2) & 6.88 $\pm$ 0.04 (3) &                    \\
1017 &  3.188 $\pm$ 0.035 (1) & 3.23 $\pm$ 0.05 (2) & 3.10 $\pm$ 0.02 (3) &                    \\
1409 &  2.671 $\pm$ 0.032 (1) & 2.67 $\pm$ 0.04 (2) &                     & 2.57 $\pm$ 0.06 (4) \\
2473 &  4.703 $\pm$ 0.047 (1) & 4.78 $\pm$ 0.07 (2) & 4.73 $\pm$ 0.03 (3) &                    \\
3249 &  5.238 $\pm$ 0.069 (1) & 5.20 $\pm$ 0.07 (2) & 5.03 $\pm$ 0.04 (3) &                    \\
4932 &  3.283 $\pm$ 0.033 (1) & 3.28 $\pm$ 0.05 (2) & 3.17 $\pm$ 0.03 (3) &                    \\
5681 &  2.764 $\pm$ 0.030 (1) & 2.75 $\pm$ 0.04 (2) & 2.74 $\pm$ 0.03 (3) & 2.71 $\pm$ 0.06 (4) \\
6220 &  2.624 $\pm$ 0.034 (1) & 2.64 $\pm$ 0.04 (2) & 2.42 $\pm$ 0.07 (3) &                    \\
6418 &  5.275 $\pm$ 0.067 (1) & 5.27 $\pm$ 0.07 (2) & 5.20 $\pm$ 0.03 (3) &                    \\
7525 &  7.271 $\pm$ 0.073 (1) & 7.24 $\pm$ 0.10 (2) & 7.08 $\pm$ 0.05 (3) &                    \\
7796 &  3.017 $\pm$ 0.030 (1) & 3.03 $\pm$ 0.04 (2) & 3.02 $\pm$ 0.08 (3) &                    \\
8079 &  5.787 $\pm$ 0.058 (1) & 5.80 $\pm$ 0.13 (2) & 5.56 $\pm$ 0.04 (3) &                    \\
8414 &  3.237 $\pm$ 0.057 (1) & 3.20 $\pm$ 0.05 (2) & 3.08 $\pm$ 0.03 (3) &                    \\
8684 &  2.496 $\pm$ 0.040 (1) & 2.49 $\pm$ 0.04 (2) & 2.50 $\pm$ 0.08 (3) & 2.60 $\pm$ 0.06 (4) \\ 
\hline
\end{tabular}

References. (1) Mozurkewich et al. (2003); (2) Nordgren et al. (2001);
(3) Nordgren et al. (1999); (4) van Belle et al. (1999).
\label{TAB-II}
\end{table}

\section{The data}

The principal sources of {\it limb-darkened} angular diameters are the
recent papers by Mozurkewich et al. (2003), Nordgren et al. (1999,
2001), van Belle et al. (1999), Kervella et al. (2003, 2004) and
Wittkowski et al. (2004). From the compilation by Richichi \&
Percheron (2002), those stars were selected which have a 
{\it limb-darkened} angular diameter determined, and with a relative
accuracy of better than 3\%. This selection is based on the typical
accuracy that can be achieved with modern instrumentation and was
selected to weed out most of the less accurate, older, data in this
compilation. The stars already in the other references were removed,
as well as the carbon- and S-stars. In total this added another 27
objects. The final list contains 221 (different) objects, listed in
Table~\ref{TAB-LIST}. In case of multiple angular diameter
determinations the one with the smallest relative error on the
limb-darkened angular diameter was retained.

For fifteen objects there are three or four independent determinations
available and its interesting to compare them as e.g. slightly
different approaches are used in the literature to proceed from
uniform-disk to limb-darkened disk angular parameters.
Table~\ref{TAB-II} lists those stars and the measurements, and
Figure~\ref{II} plots the values from Nordgren et al. (2001) against
those in Mozurkewich et al. (2003). The agreement is extremely good,
and the rms in the fit of 0.039 indicates that the quoted errorbars in
the determinations are realistic.

Apart from the angular diameters, one needs apparent magnitudes and
reddening to apply the SB method. Most of the photometric data come
from consulting the SIMBAD database. In particular, IRAS 12 and 25
$\mu$m fluxes, and $BVRIJK$ magnitudes were collected from the entries
marked (in order of preference), ``JP11'', ``UBV'', ``IRC''. In
addition, the galactic coordinates, spectral type, and the \HP\
parallax was retrieved, as well as the most recent listed value for
the metallicity, available under the item marked ``Fe\_H''.  To
collect more photometric data, the Gezari et al. (1999) catalog was
checked for infrared data, and the SIMBAD based list of references was
checked for papers that, based on the title alone, might contain
additional photometry. For 43 stars only a $K$-mag from the IRC-survey
is available, and it was investigated if there is a systematic
difference between Johnson $K$ and IRC $K$. Based on 66 objects (minus
3 outliers) a difference $K$(Johnson - IRC) = 0.017 $\pm$ 0.045 was
established, and therefore no correction to the IRC magnitudes was
applied. It may appear attractive to include \M\ photometry for all
stars--for reasons of uniformity--or at least for those stars without
infrared photometry sofar. However, the stars in the sample are so
bright that no \M\ photometry could be obtained in the standard way. 
Figure~\ref{JKJK} shows for  119 stars with $JK$ photometry on
the Johnson system and \M\ photometry with a quoted errorbar in both
filters of less than 0.3 mag the correlation between 
$(J-K)_{\rm Johnson}$ and $(J-K)_{\rm 2MASS}$ and 
$K_{\rm Johnson}-K_{\rm 2MASS} $ and $(J-K)_{\rm 2MASS}$. The rms in
the fits are about 0.1 mag, and so applying these relations would
cause unwanted scatter, and therefore \M\ data has not been used. 
It should be noted that some ``natural'' scatter is present in the
adopted photometry because e.g. of non-simultaneous photometry or low
level variability.

The finally adopted magnitudes are listed in Table~\ref{TAB-LIST}, and are
predominantly on the Johnson system. The IRAS flux densities were
converted to magnitudes assuming zero points of 41.0 and 9.49 Jy at 12
and 25 $\mu$m respectively.

The reddening was estimated using the reddening model by Arenou et al. (1992), 
which returns the reddening estimate (with error) based on galactic
coordinates and distance. The distance used was simply based on the
\HP\ parallax, and when no parallax was available a distance of 1 kpc
was assigned. As discussed later, none of these assumptions is
critical to the final results of this paper.  The adopted values for
$A_{\rm V}$ and its error are listed in Table~\ref{TAB-LIST}. The
relative reddening $A_{\lambda}/A_{\rm V}$ are adopted to be 1.33,
0.80, 0.49, 0.27, 0.12, 0.034 and 0.018 in $BRIJK$[12][25],
respectively, following Draine (2003).

In what follows we will the use the following terminology: giants are
objects which have ``{\sc iii}'' in their spectral type listed in
Table~\ref{TAB-LIST} (i.e. includes stars with spectral type {\sc
ii-iii}), dwarfs are objects of luminosity class {\sc v}, sub-giants
are objects of luminosity class {\sc iv}, supergiants are objects of
luminosity class {\sc I}, and luminous giants are objects of
luminosity class {\sc ii}.  Three objects have no luminosity class
assigned in SIMBAD and for those the absolute $V$-magnitude and
$(B-V)_0$ were determined from the parallax and reddening, and
compared to the tables of Strai\v{z}ys \& Kuriliene (1981) and
Schmidt-Kaler (1982). Based on this, HR 6208 ($M_{\rm V} = 0.3 \pm
0.2$) was assigned luminosity class {\sc iii}, and HR 8555 ($M_{\rm V}
= -1.5 \pm 0.4$) and IRC +40 337 ($M_{\rm V} = -1.6 \pm 0.8$) were
assigned luminosity class {\sc ii}.

\section{Basic relations}

A surface-brightness relation can be defined as follows (see van Belle 1999):
\begin{equation}
\theta_o = \theta \times 10^{(m_1/5)},
\end{equation}
where $\theta$ is the (limb-darkened) angular diameter (in mas), and
$m_1$ a de-reddened magnitude (for example, $V$). The logarithm of
this quantity is plotted against a de-reddened colour (for example,
$(V-K)_0$), 
\begin{equation}
\log \theta_0 = a \times (m_2 - m_3) + b, 
\end{equation}
and a linear fit is performed.
An equivalent form is to use the quantity (see, e.g., Nord02)
\begin{equation}
F_{m_1} = 4.2207 -0.1 \,m_1 -0.5 \log \theta,
\end{equation}
and to plot this against a colour; 
\begin{equation}
F_{m_1} = a^\prime \times (m_2 - m_3) + b^\prime.
\end{equation}
The former mathematical formulation will be used in this paper, but it is trivial to show that,
\begin{equation}
a = -2 \; a^\prime {\;\;\;\; \rm and \;\;\;\;} b = 2 \;( 4.2207 - b^\prime).
\end{equation}
In the analysis below, the error on $\theta_0$ includes the error on $\theta$, an arbitrary
but representative error of 0.01 mag in the photometry, and the error
on the reddening.

\section{Results}

%awk '{print $4}' barnes_evans_simbad.dat | grep M | grep -v V 
%awk '{if ($6>-5. && $10>-5. && $11> -5. && $12> -5 ) print $4}' barnes_evans_simbad.dat 
%awk '{if ($6<-5. || $10<-5. || $11< -5. || $12< -5 ) print $6, $10, $11, $12}' barnes_evans_simbad.dat 

The assumption in the above relations is that, after correction for
interstellar reddening, the colours represent the photospheric
colours. As a large fraction of the stars among the sample of 221 are
M giants (59) or supergiants (7) one might need to consider the
influence of mass loss on the colours.

Figure \ref{Fig-CC} shows in the top panel the IRAS [12-25] colour
versus $(V-K)$ with different luminosity classes indicated by
different symbols for the 195 stars for which these four colours are
available. The large majority of stars scatter around [12-25]
$\approx$ 0, as expected, but for redder $(V-K)$ there are nine stars
with a clear mid-IR excess. They are HR 681 (M7 IIIe), HR 1845 (M2
Iab), HR 2061 (M1 Iab), HR 6146 (M6 III), HR 8316 (M2 Iae), IRC +20282
(M5 II-III), IRC +30115 (M1.5 Ia), IRC +30257 (M7.5 III), IRC +30412
(M3 Ib), and are excluded in the fitting described below. For
comparison, the lower panel of the figure presents the colour-colour
diagram excluding all stars of spectral type M (giants and
supergiants). Statistically, 9 of the 63 M-giants and supergiants in
this sub-sample have a mid-IR excess. Of the 26 stars without $V$ or
$K$ or IRAS photometry there are three additional M-stars, and
therefore, statistically, none is expected to have an mid-IR excess.

\begin{figure}
\includegraphics[width=85mm]{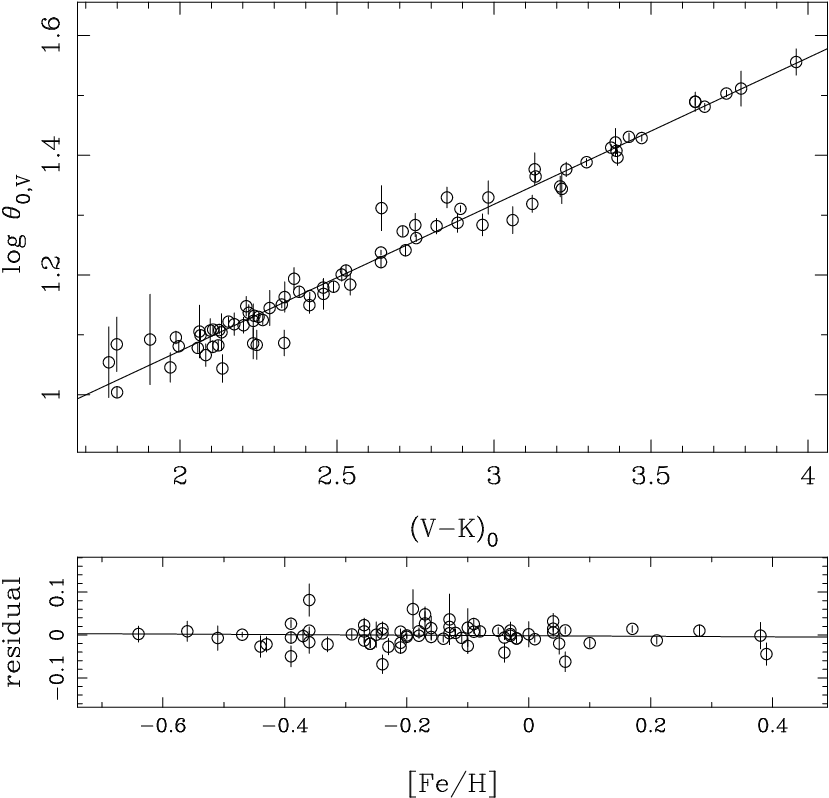}
\caption[]{
Preferred $V$-band surface-brightness relation versus colour $(V-K)_0$ for non-M giants.
Symbols as in Figure~\ref{Fig-CC}.
In the bottom panel the residuals are plotted versus metallicity.
}
\label{Fig-first}
\end{figure}

\begin{figure}
\includegraphics[width=85mm]{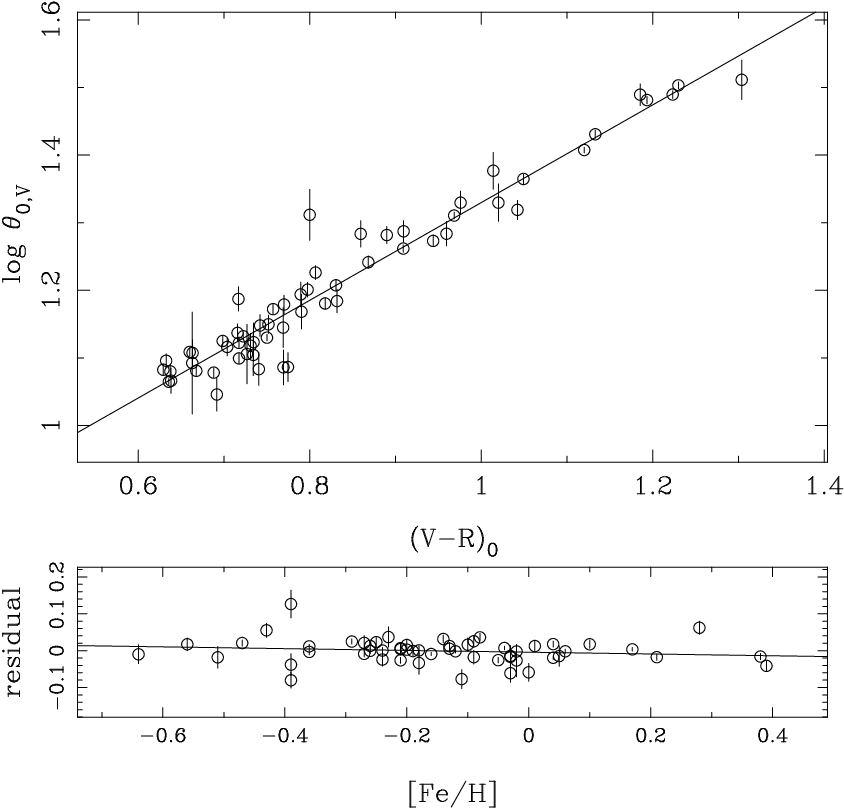}
\caption[]{
Preferred $V$-band surface-brightness relation versus colour $(V-R)_0$ for non-M giants.
The residuals correlate at the $1\sigma$ level with metallicity: residual $= (-0.024 \pm 0.021)$ [Fe/H] + $(-0.0041 \pm 0.0050)$.
}
\label{Fig-V-VR1}
\end{figure}

\begin{figure}
\includegraphics[width=85mm]{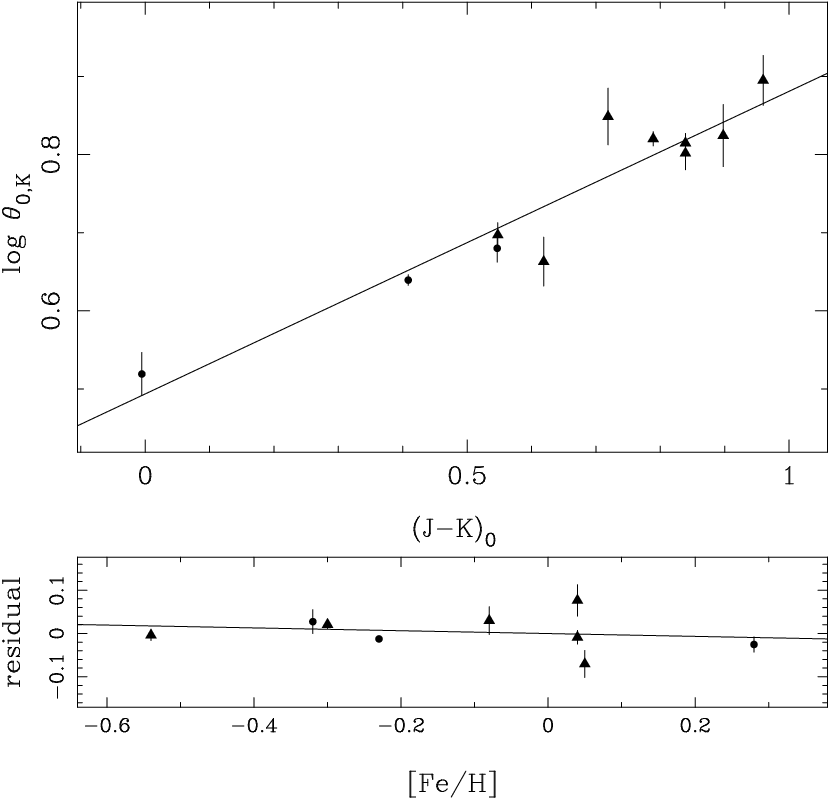}
\caption[]{
Final $K$-band surface-brightness relation versus colour $(J-K)_0$ for dwarfs and sub-giants.
The residuals correlate weakly with metallicity: residual $= (-0.033 \pm 0.061)$ [Fe/H] + $(0.000 \pm 0.016)$.
}
\label{Fig-last}
\end{figure}

Following the literature, the following SB relations are considered:
$V$ versus $(V-R)$, $V$ versus $(V-K)$, and $K$ versus $(J-K)$ (e.g.
Fouqu\'e \& Gieren 1997). The results of the fitting are listed in
Table~\ref{TAB-RES} and shown in Figures~\ref{Fig-first}-\ref{Fig-last}. 
The table lists the values of the fitting coefficients $a$ and $b$,
the number of stars considered, the rms in the fit, the colour range
over which the fit was performed, the luminosity class, and some
additional remarks. In the bottom panel of all figures the residual is
plotted against metallicity, and the fit is explicitly given when
significant. The possible significance of this will be discussed in Section~5.

The fit was performed using a least-squares algorithm. Outliers were
identified and removed, and the fit repeated. Outliers were again
removed and the final fit made. An object was considered an outlier if
the distance between a data point and the fitted line was more than 5
times either the error bar in the data point, or the rms in the fit. 
Table~\ref{TAB-RES} also shows some results when the $\sigma$-clipping
is more stringent. The rms in the fit are obviously reduced but not by
much, and the coefficients change within their error bars.

Table~\ref{TAB-RES} also includes the result when not considering the
M-giants. In that case the colour range over which the relation is
applicable becomes smaller, but the relations might be more
appropriate for Cepheids for which these relations are often used. The
number of stars in the sample is reduced and hence the formal error on
the coefficients becomes slightly larger. The rms in the fits are not
appreciably reduced. Considering only the M-giants results in a
relation which is significantly different from the relation for all
giants, and all non-M giants, as suggested already by DB93.

To investigate the influence of reddening a set of 1000 Monte Carlo
simulations was performed where each time the reddening was replaced
by a value drawn from a Gaussian distribution with mean the adopted
reddening and sigma the error therein. The 23rd and 977th ordered
value correspond to $\pm$2$\sigma$, and in the case of the giants
(first entry in Table~\ref{TAB-RES}), the value and the error on $a$
and $b$ are (0.2368 $\pm$ 0.0005, 0.6080 $\pm$ 0.0012), (0.7279 $\pm$
0.0011, 0.6076 $\pm$ 0.0010) and (0.1607 $\pm$ 0.0038, 0.5905 $\pm$
0.0030), for the case $V$ versus $(V-K)$, $V$ versus $(V-R)$ and $K$
versus $(J-K)$, respectively. The conclusion is that, formally, the
influence of reddening is smallest in the case $V$ versus
$(V-K)$. However, in all three cases the error in the reddening is
insignificant compared to the fit error.

\begin{figure}
\includegraphics[width=85mm]{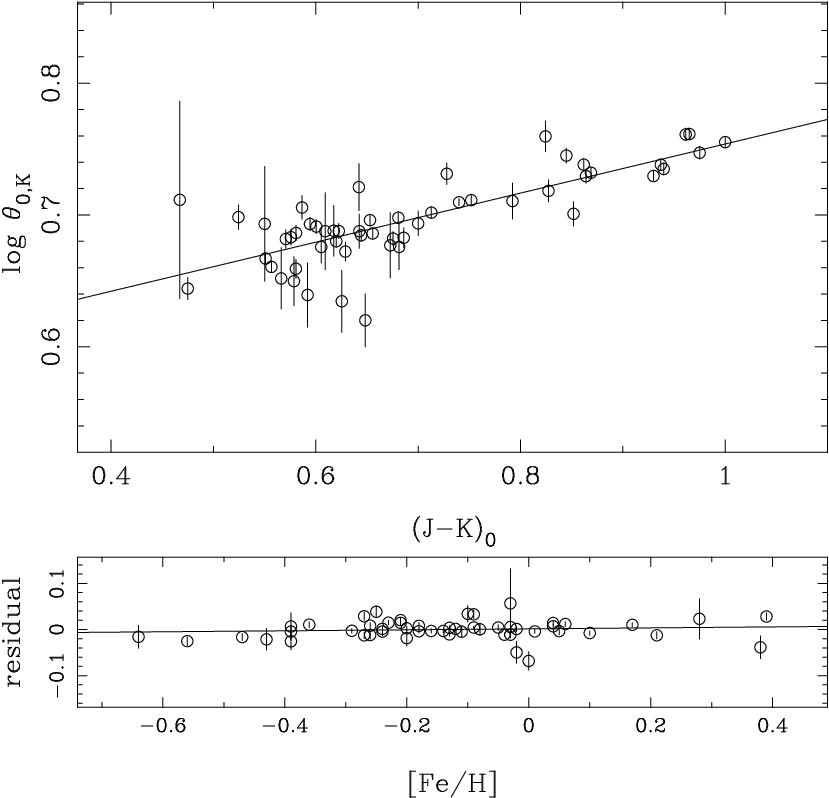}
\caption[]{
Preferred $K$-band surface-brightness relation versus colour $(J-K)_0$ for non-M giants.
}
\label{Fig-J-JK}
\end{figure}

\begin{figure}
\includegraphics[width=85mm]{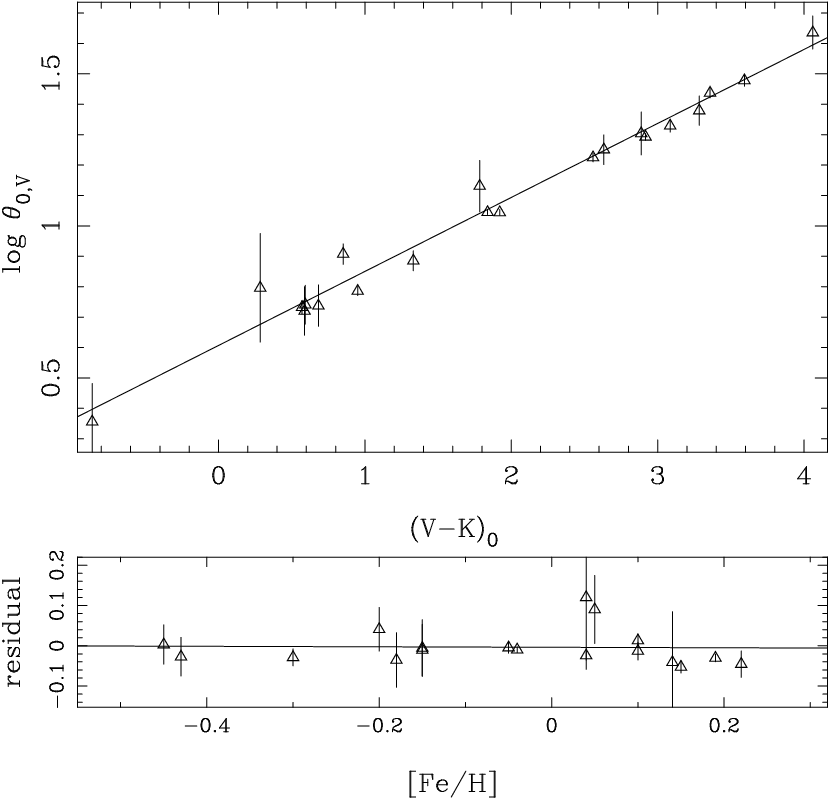}
\caption[]{
Final $V$-band surface-brightness relation versus colour $(V-K)_0$ for supergiants.
}
\label{Fig-V-VK}
\end{figure}

\begin{figure}
\includegraphics[width=85mm]{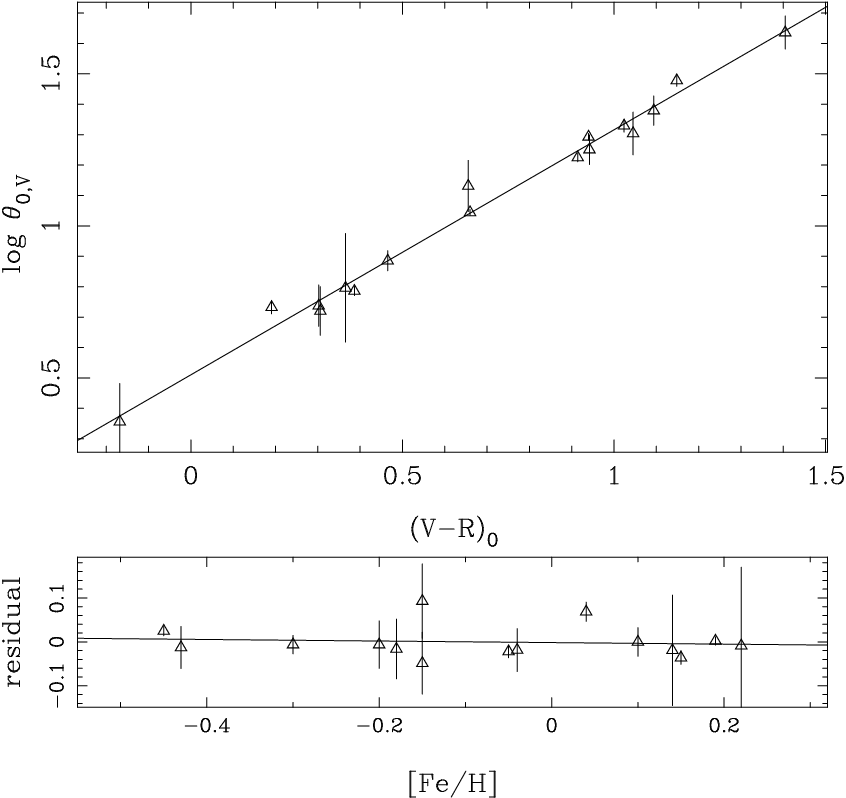}
\caption[]{
Final $V$-band surface-brightness relation versus colour $(V-R)_0$ for supergiants.
}
\label{Fig-V-VR}
\end{figure}

\begin{figure}
\includegraphics[width=85mm]{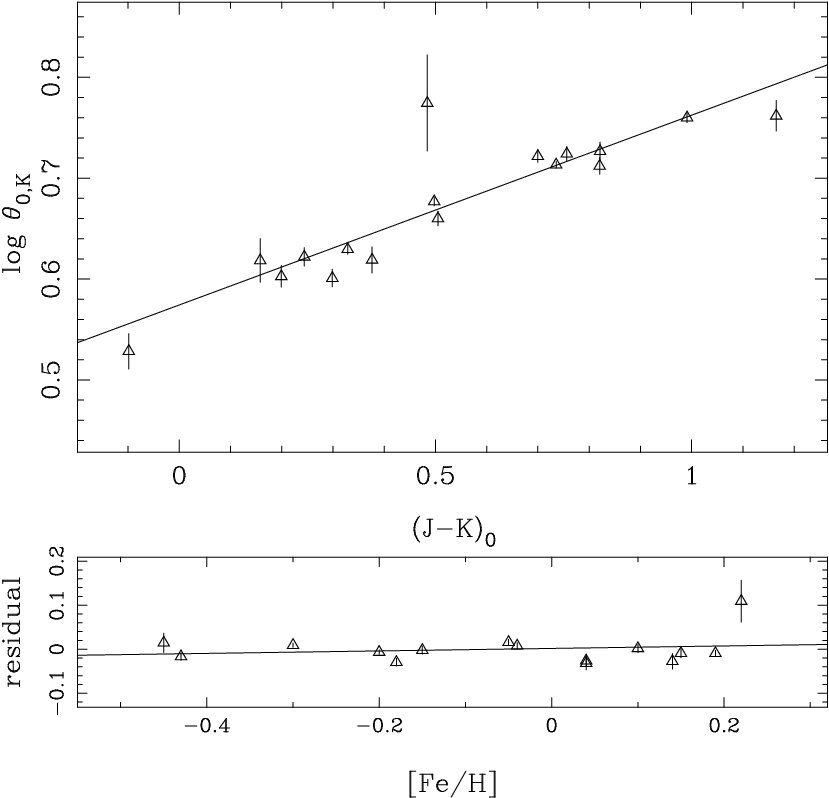}
\caption[]{
Final $K$-band surface-brightness relation versus colour $(J-K)_0$ for supergiants. 
The residuals correlate weakly with metallicity: residual $= (0.028 \pm 0.042)$ [Fe/H] + $(0.002 \pm 0.009)$.
}
\label{Fig-J-JK}
\end{figure}

\begin{figure}
\includegraphics[width=85mm]{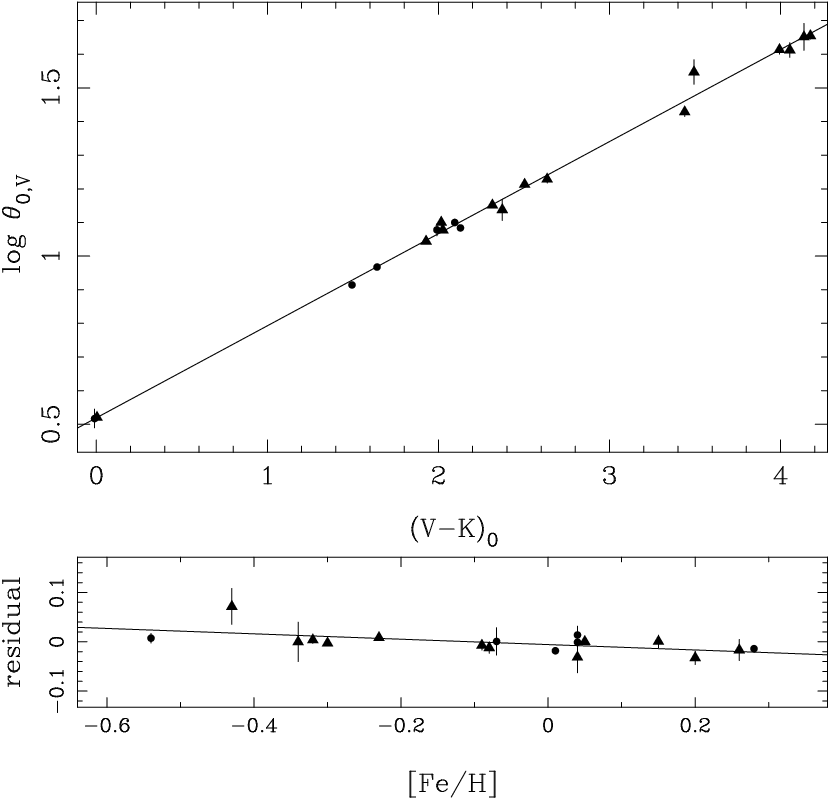}
\caption[]{
Final $V$-band surface-brightness relation versus colour $(V-K)_0$ for  dwarfs and sub-giants.
The residuals correlate significantly with metallicity: residual $= (-0.054 \pm 0.019)$ [Fe/H] + $(-0.006 \pm 0.005)$.
}
\label{Fig-V-VK}
\end{figure}

\begin{figure}
\includegraphics[width=85mm]{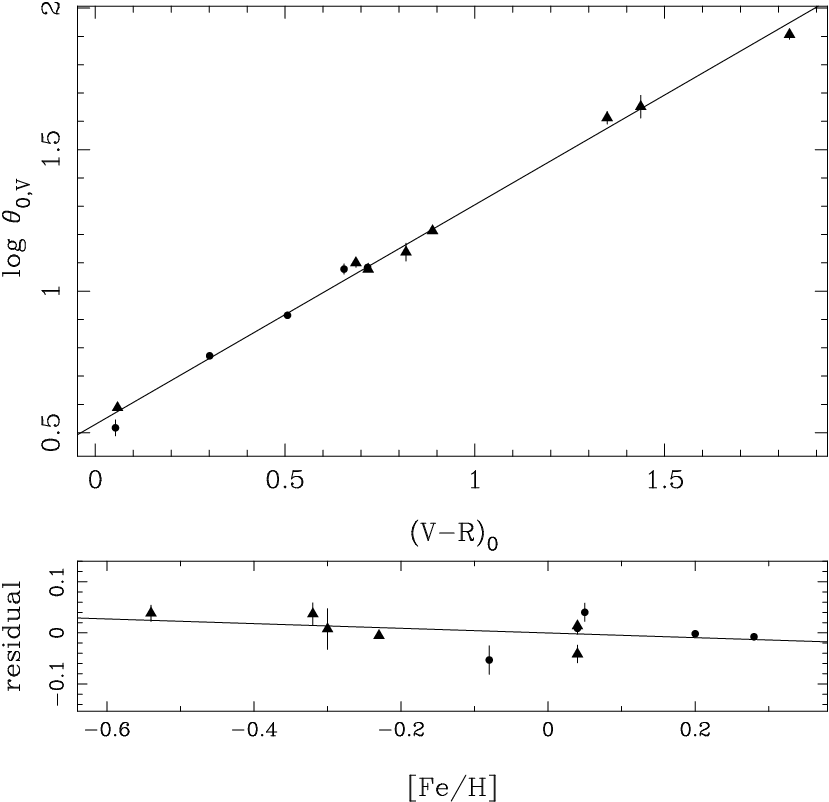}
\caption[]{
Final $V$-band surface-brightness relation versus colour $(V-R)_0$ for  dwarfs and sub-giants.
The residuals correlate at the $1\sigma$ level with metallicity: residual $= (-0.046 \pm 0.038)$ [Fe/H] + $(0.000 \pm 0.009)$.
}
\label{Fig-V-VR}
\end{figure}

\begin{table*}
\caption{SB relations derived in this work. For the giants the solutions marked by a $\bullet$ are the preferred ones.}
\begin{tabular}{ccccrlrllll} \hline
$(m_1)$ &     $a$   & $(m_2 - m_3)$ &      $b$       & $N$ &  rms  &  Range     & Remark  \\
\hline
 $V$ & 0.237 $\pm$ 0.003 & (V-K) & 0.608 $\pm$ 0.009 & 122 & 0.027 & 1.7-5.0 & giants, cut at 5$\sigma$   \\% 0.0018 0.016  0.00074 0.0038 N=82
 $V$ & 0.239 $\pm$ 0.002 & (V-K) & 0.597 $\pm$ 0.008 & 114 & 0.024 & 1.7-5.0 & giants, cut at 4$\sigma$ \\% -0.011 0.013 -0.0019 0.0030 N=82  
 $V$ & 0.236 $\pm$ 0.002 & (V-K) & 0.608 $\pm$ 0.007 &  99 & 0.020 & 1.7-5.0 & giants, cut at 3$\sigma$ \\% 0.012 0.011  0.0005 0.0027 N=76
 $V$ & 0.242 $\pm$ 0.006 & (V-K) & 0.595 $\pm$ 0.016 &  82 & 0.029 & 1.7-4.0 & giants, no M-giants, cut at 5$\sigma$ \\% 0.0019 0.017  0.0011 0.0043 N=76 
$\bullet$ $V$ & 0.245 $\pm$ 0.005 & (V-K) & 0.584 $\pm$ 0.014 &  76 & 0.024 & 1.7-4.0 & giants, no M-giants, cut at 4$\sigma$ \\% -0.0067 0.014 -0.0018 0.0037 N=70 
 $V$ & 0.241 $\pm$ 0.004 & (V-K) & 0.595 $\pm$ 0.012 &  70 & 0.020 & 1.7-4.0 & giants, no M-giants, cut at 3$\sigma$ \\%-0.0055 0.012  -0.0013 0.0029 N=67
 $V$ & 0.237 $\pm$ 0.007 & (V-K) & 0.610 $\pm$ 0.028 &  40 & 0.021 & 3.2-6.1 & M-giants, cut at 4$\sigma$ \\% 0.032 0.047  0.0038 0.0080 N=6 
\\
 $V$ & 0.728 $\pm$ 0.012 & (V-R) & 0.608 $\pm$ 0.012 &  75 & 0.031 & 0.6-1.8 & giants \\% -0.011 0.020  -0.0028 0.0050   N=61
 $V$ & 0.735 $\pm$ 0.011 & (V-R) & 0.601 $\pm$ 0.010 &  70 & 0.026 & 0.6-1.8 & giants, cut at 4$\sigma$ \\% 0.0056 0.016  -0.0004 0.0040   56
 $V$ & 0.724 $\pm$ 0.009 & (V-R) & 0.617 $\pm$ 0.009 &  60 & 0.020 & 0.6-1.8 & giants, cut at 3$\sigma$ \\% -0.024 0.014  -0.0025 0.0033   48
 $V$ & 0.716 $\pm$ 0.022 & (V-R) & 0.613 $\pm$ 0.020 &  65 & 0.032 & 0.6-1.3 & giants, no M-giants, cut at 5$\sigma$ \\% -0.021 0.020  -0.0023 0.0052   62
$\bullet$ $V$ & 0.723 $\pm$ 0.025 & (V-R) & 0.607 $\pm$ 0.021 &  59 & 0.032 & 0.6-1.3 & giants, no M-giants, cut at 4$\sigma$ \\% -0.024 0.021  -0.0041 0.0050   56
 $V$ & 0.730 $\pm$ 0.017 & (V-R) & 0.598 $\pm$ 0.015 &  47 & 0.021 & 0.6-1.3 & giants, no M-giants, cut at 3$\sigma$ \\% -0.0028 0.015  -0.0000 0.0038   45
 $V$ & 0.528 $\pm$ 0.045 & (V-R) & 0.912 $\pm$ 0.065 &   9 & 0.017 & 1.2-1.7 & M-giants, cut at 4$\sigma$ \\% -0.124 inf  0.0084 inf  N=2 
\\
 $K$ & 0.161 $\pm$ 0.015 & (J-K) & 0.591 $\pm$ 0.012 &  81 & 0.027 & 0.5-1.3 & giants \\%-0.038 0.017 -0.0052 0.0040   63
 $K$ & 0.170 $\pm$ 0.012 & (J-K) & 0.579 $\pm$ 0.010 &  70 & 0.021 & 0.5-1.3 & giants, cut at 4$\sigma$ \\%-0.0022 0.015 -0.0007 0.0035 N=55
 $K$ & 0.196 $\pm$ 0.011 & (J-K) & 0.560 $\pm$ 0.009 &  56 & 0.015 & 0.5-1.3 & giants, cut at 3$\sigma$ \\% 0.010 0.013 0.0007 0.0029 N=43
 $K$ & 0.166 $\pm$ 0.026 & (J-K) & 0.585 $\pm$ 0.019 &  57 & 0.027 & 0.5-1.0 & giants, no M-giants, cut at 5$\sigma$ \\ % 0.0048 0.018 0.0009 0.0044   55
$\bullet$ $K$ & 0.186 $\pm$ 0.021 & (J-K) & 0.568 $\pm$ 0.015 &  54 & 0.021 & 0.5-1.0 & giants, no M-giants, cut at 4$\sigma$ \\% 0.0011 0.014 0.0014 0.0035   52
 $K$ & 0.186 $\pm$ 0.018 & (J-K) & 0.564 $\pm$ 0.012 &  42 & 0.015 & 0.5-1.0 & giants, no M-giants, cut at 3$\sigma$ \\% -0.011 0.012 -0.0014 0.0027   40
 $K$ & 0.081 $\pm$ 0.042 & (J-K) & 0.682 $\pm$ 0.044 &  20 & 0.021 & 0.7-1.3 & M-giants, cut at 4$\sigma$ \\% 0.026 0.062 -0.0061 0.011   6 
\\
 $V$ & 0.223 $\pm$ 0.008 & (V-K) & 0.674 $\pm$ 0.033 &  18 & 0.046 & 1.5-6.3 & luminous giants, cut at 4$\sigma$ \\%-0.0056 0.061  0.0070 0.016 N=11
 $V$ & 0.684 $\pm$ 0.023 & (V-R) & 0.649 $\pm$ 0.025 &  12 & 0.024 & 0.7-1.8 & luminous giants, cut at 4$\sigma$ \\%-0.048 0.025  -0.0060 0.0072 N=9
 $K$ & 0.198 $\pm$ 0.018 & (J-K) & 0.576 $\pm$ 0.017 &   7 & 0.009 & 0.5-1.2 & luminous giants, cut at 4$\sigma$ \\% 0.016 0.046  -0.0023 0.0065 N=4 
\\
 $V$ & 0.240 $\pm$ 0.005 & (V-K) & 0.601 $\pm$ 0.014 &  86 & 0.026 & 1.7-4.0 & Luminosity Class {\sc ii + iii}, non-M giants, cut at 4$\sigma$ \\% -0.0030 0.014  -0.0000 0.0033 N=80
 $V$ & 0.713 $\pm$ 0.024 & (V-R) & 0.616 $\pm$ 0.020 &  68 & 0.032 & 0.6-1.3 & Luminosity Class {\sc ii + iii}, non-M giants, cut at 4$\sigma$ \\% -0.014 0.018  -0.0023 0.0047  N=65
 $K$ & 0.184 $\pm$ 0.019 & (J-K) & 0.570 $\pm$ 0.014 &  61 & 0.020 & 0.5-1.0 & Luminosity Class {\sc ii + iii}, non-M giants, cut at 4$\sigma$ \\%  0.0015 0.013  0.0002 0.0032  N=59 
\\
 $V$ & 0.243 $\pm$ 0.008 & (V-K) & 0.607 $\pm$ 0.019 &  21 & 0.046 & $-$0.9-4.1 & supergiants, cut at 4$\sigma$ \\% -0.0057 0.057 -0.0037 0.0101 N=18
 $V$ & 0.806 $\pm$ 0.023 & (V-R) & 0.510 $\pm$ 0.018 &  17 & 0.036 & $-$0.2-1.4 & supergiants, cut at 4$\sigma$ \\% -0.017 0.047  -0.0017 0.011  15
 $K$ & 0.188 $\pm$ 0.024 & (J-K) & 0.574 $\pm$ 0.015 &  17 & 0.031 & $-$0.1-1.2 & supergiants, cut at 4$\sigma$ \\%  0.028 0.042  0.0017 0.0094  15 
% $V$ & 0.239 $\pm$ 0.002 & (V-K) & 0.603 $\pm$ 0.007 & 154 & 0.033 & $-$0.9-6.7 & giants \& supergiants \\%0.0124 0.015  0.0032 0.0036 N=104
% $V$ & 0.773 $\pm$ 0.011 & (V-R) & 0.558 $\pm$ 0.011 & 106 & 0.040 & $-$0.2-2.0 & giants \& supergiants \\%-0.021 0.021 -0.0043 0.0050
% $K$ & 0.153 $\pm$ 0.012 & (J-K) & 0.597 $\pm$ 0.009 &  98 & 0.030 & $-$0.1-1.2 & giants \& supergiants \\% -0.036 0.016 -0.0035 0.0038
\\
 $V$ & 0.274 $\pm$ 0.004 & (V-K) & 0.519 $\pm$ 0.012 &   20 & 0.022 & $-$0.1-4.2 & dwarfs \& sub-giants, cut at 4$\sigma$ \\% -0.054 0.019  -0.0056 0.0046 N=18
 $V$ & 0.776 $\pm$ 0.017 & (V-R) & 0.529 $\pm$ 0.016 &   13 & 0.028 &    0.1-1.8 & dwarfs \& sub-giants, cut at 4$\sigma$ \\% -0.046 0.038   0.00014 0.0094  11
 $K$ & 0.387 $\pm$ 0.045 & (J-K) & 0.494 $\pm$ 0.032 &   11 & 0.036 &    0.0-1.0 & dwarfs \& sub-giants, cut at 4$\sigma$ \\% -0.033 0.061   0.00000 0.016    9

\hline
\end{tabular}
\label{TAB-RES}
\end{table*}

\section{Discussion}

Surface brightness relations have been derived for a large number of
stars making use of the large datasets on angular diameter
measurements for ``normal'' stars that have become available recently.
In the present section the results will be discussed, and compared to
earlier work. For convenience, SB relations reported in the literature
have been compiled in Table~\ref{TAB-PREV} following the format of
Table~\ref{TAB-RES} (usually $a^\prime$ and $b^\prime$ are quoted, and
they have been converted to $a$ and $b$ by me. The rms is the one
quoted for the original fit--which therefore almost always refer to Eq.~(4)--,
and its value should be multiplied by a factor of 2 to compare to the
rms values listed in Table~\ref{TAB-RES} !)

Recently, Kervella et al. (2004) presented SB-relations for stars of
luminosity class {\sc iv + v} of the type $m_1$ versus ($m_1 - m_2)$
for all possible combinations for magnitudes $UBVRIJHKL$. The
agreement is perfect in the case $V - (V-K)$ but this should not be
surprising as essentially the same angular diameter and photometric
data has been used. For the other 2 relations the fits not in good
agreement.  In those two cases the clipping at 4$\sigma$ has removed a
substantial number of objects in the present paper, while Kervella et
al. appear not to have used clipping but remark that the relation are
in fact non-linear. It appears that for dwarfs the $V - (V-K)$
relation is to be preferred.

For the supergiants a comparison can be made with FG97. For the
$V-(V-K)$ and $V-(V-R)$ relation the sample considered here is almost
double that of FG97, and extends to bluer colours. The error on the
coefficients is correspondingly smaller, yet the rms is in fact
slightly larger. For typical colours $(V-K)_0 = 2.0$, $(V-R)_0 = 0.5$,
and $(J-K)_0 = 0.5$ the relation in the present paper predicts $\log
\theta_0$ of 1.093, 0.913 and 0.668, respectively, while FG97 predict
1.089, 0.947 and 0.666, respectively. There are some differences but
not systematically so it appears. The relations in the present paper
should be preferred because they are based on a larger number of stars.

Probably of most interest is the relation for giants. This is also the
class of stars for which most data is available. A first important
conclusion is that the relation for the M-giants is different from
that for all giants and for all non-M giants. This was first hinted at
by DB93 based on few stars only, but is confirmed here based on a much
larger sample. Like him it is found that $a$ is smaller and $b$ larger
for M-giants than for the non-M giants, in particular for the
$V-(V-R)$ and $K-(J-K)$ relations at a level of 3-5$\sigma$. This
makes a comparison with FG97 and Nord02 difficult as they did not
exclude M-giants in deriving their SB-relations.  For typical colours
$(V-K)_0 = 2.5$, $(V-R)_0 = 0.9$, and $(J-K)_0 = 0.7$ the preferred
relation for non-M giants in the present paper predicts $\log
\theta_0$ of 1.197, 1.258 and 0.698, respectively, while FG97 (Nord02)
predict for (all) giants 1.201 (1.188), 1.272 (1.248), 0.701
(0.690). For all three relations the sizes are slightly smaller than
predicted by FG97--which is consistent with the fact that it is found
in the present paper that the M-giants are bigger than the non-M
giants--but smaller than the values listed in Nord02 which {\it
includes} M-giants.

As was noted earlier in the literature, including more stars does not
lead to a smaller rms. Considering giants of all types, the rms in,
for example, the $V-(V-K)$ relation is 0.016 (based on 10 stars in
FG97), 0.022 (based on 57 stars in Nord02) and 0.027 (based on 122
stars in the present paper). It is shown that the SB relations do not
significantly depend on metallicity, and hence this scatter is not due
to that parameter. Specifically for the giants, only the residuals in
the $V-(V-R)$ relation correlate at the 1$\sigma$ level with
metallicity (see caption of Figure~\ref{Fig-V-VR1}).

\begin{figure}
\includegraphics[width=85mm]{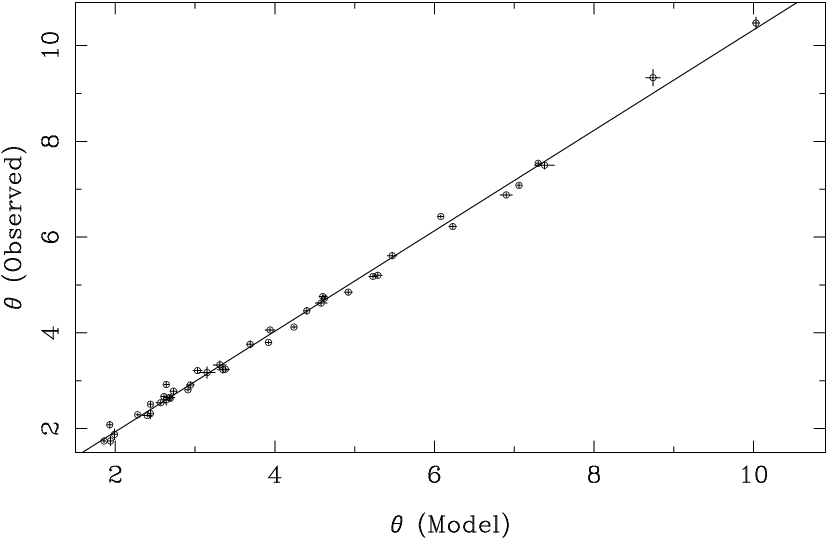}
\caption[]{
The adopted observed limb-darkened angular diameters (in mas) plotted
against the radioametrically predicted values from Cohen et al. (1999). 
The line is a least-squares fit to the 42 data points plotted:
%(3 outliers at the level of 4$\sigma$ have been removed: HR 2012, 4546 and 7942): 
$\theta_{\rm observed} = (1.049 \pm 0.010) \;
\theta_{\rm model} + (-0.166 \pm 0.048)$ (rms = 0.130).
}
\label{MODEL}
\end{figure}

Cohen et al. (1999) predict angular diameters for more than 400 giants
in the spectral range G9.5 to M0 by fitting model atmospheres to
absolute flux-calibrated broad-band photometry. Already in their paper
they compared the predicted values to observed values, for about 20
stars.  Figure~\ref{MODEL} shows the same comparison but now for a
sample twice in size. A least-square fit was made and outliers at the
4$\sigma$ level removed (HR 2012, 4546). A new fit was made, and one
additional outlier (HR 7942) was removed. The final fit is:
$\theta_{\rm observed} = (1.049 \pm 0.010) \; \theta_{\rm model} +
(-0.166 \pm 0.048)$, with an rms of 0.130. The slope is in the same
sense as found by Cohen et al. (1.013 $\pm$ 0.008), i.e. slightly
above unity, and found to be slightly steeper now, while the zero
point is now significantly below zero, while the value in Cohen et
al. was consistent with zero (0.035 $\pm$ 0.073). No immediate
explanation is offered but it is simply noticed that the rms in the
fit is larger than expected based on the assigned errorbars. As
independent observations seem to be in very good agreement (see
Figure~\ref{II}) any possible problem seems more likely to be related
to the models rather than the observations. In addition, any possible
problem appears to be for the larger objects. Selecting only stars
smaller than a certain cut-of and monitoring the significance of the
derived slope and zero point it is found that for $\theta <$ 6 mas the
expected 1-to-1 relation is found $\theta_{\rm observed} = (1.009 \pm
0.019) \; \theta_{\rm model} + (-0.028 \pm 0.066)$, although the rms
is only reduced to 0.117 mas.

A final point that has to be addressed is the practice in the
literature (FG97 and Nord02) to impose a common zero point of SB
relations. In practice, the same three relations used here
(i.e. $V-(V-K)$, $V-(V-R)$ and $K-(J-K)$) are first derived
independently. Then the zero point is averaged. Finally new slopes are
derived for the three relations keeping the zero point fixed to this
averaged value. The argument given is along the line that ``For a star
of spectral type A0, where $(V-R)_0 = (V-K)_0 = (J-K)_0 = 0$, each of
the relations should yield a common surface brightness'' (Nord02).
Life is not that simple. Table~\ref{TAB-COL} lists the (theoretical)
colours of stars at various effective temperatures and gravities very
near the colour $(V-K) = 0$ (from Pietrinferni et al. 2004). It can be
remarked that indeed $(J-K)$ is in all cases also very near zero, but
$(V-R)$ is near zero only for main-sequence stars but not for lower
gravities. Imposing a common zero point based on simple averaging
individual zero points is therefore not allowed for giants or
supergiants. The value derived in such a way is biased (and this will
also influence the then newly calculated slopes) if the common zero
point is based on the zero point from the $V-(V-R)$ relation, like is
the case in FG97 and Nord02.

\begin{table*}
\setlength{\tabcolsep}{0.9mm}
\caption{Previous SB relations}
\begin{tabular}{ccccrllccll} \hline
 $(m_1)$ &    $a^\prime$      & $(m_2 - m_3)$ &    $b^\prime$     & $N$ &  rms$^{(1)}$ &  Range    & $a$ & $b$ & Remark & Reference\\
\hline
  $V$   & $-0.123 \pm$ 0.002 & (V-K) & 3.934 $\pm$ 0.005 & 57  & 0.011 & 0.7-4.1   & $0.246 \pm$ 0.004 & 0.573 $\pm$ 0.010 & giants & Nord02 \\
  $V$   & $-0.365 \pm$ 0.009 & (V-R) & 3.925 $\pm$ 0.009 & 57  & 0.016 & 0.2-1.3   & $0.730 \pm$ 0.018 & 0.591 $\pm$ 0.018 & giants & Nord02 \\
  $K$   & $-0.095 \pm$ 0.007 & (J-K) & 3.942 $\pm$ 0.006 & 57  & 0.011 & 0.18-1.1  & $0.190 \pm$ 0.014 & 0.557 $\pm$ 0.012 & giants & Nord02 \\
  $V$   & $-0.134 \pm$ 0.005 & (V-K) & 3.956 $\pm$ 0.011 & 59  & 0.026 & 0.9-2.0   & $0.268 \pm$ 0.010 & 0.529 $\pm$ 0.022 & cepheids & Nord02 \\
  $V$   & $-0.364 \pm$ 0.011 & (V-R) & 3.939 $\pm$ 0.006 & 59  & 0.026 & 0.3-0.7   & $0.728 \pm$ 0.022 & 0.563 $\pm$ 0.012 & cepheids & Nord02 \\
  $K$   & $-0.080 \pm$ 0.021 & (J-K) & 3.934 $\pm$ 0.009 & 59  & 0.026 & 0.2-0.5   & $0.160 \pm$ 0.042 & 0.573 $\pm$ 0.018 & cepheids & Nord02 \\
  $V$   & $-0.125 \pm$ 0.003 & (V-K) & 3.941 $\pm$ 0.004 & 59  &       &           & $0.250 \pm$ 0.006 & 0.559 $\pm$ 0.008 & cepheids & Nord02, forced common ZP \\
  $V$   & $-0.368 \pm$ 0.007 & (V-R) & 3.941 $\pm$ 0.004 & 59  &       &           & $0.736 \pm$ 0.014 & 0.559 $\pm$ 0.008 & cepheids & Nord02, forced common ZP \\
  $K$   & $-0.096 \pm$ 0.010 & (J-K) & 3.941 $\pm$ 0.004 & 59  &       &           & $0.192 \pm$ 0.020 & 0.559 $\pm$ 0.008 & cepheids & Nord02, forced common ZP \\
  $V$   & $-0.124 \pm$ 0.004 & (V-K) & 3.930 $\pm$ 0.012 & 10  & 0.008 & 2.22-4.11 & $0.248 \pm$ 0.008 & 0.581 $\pm$ 0.024 & giants & FG97 \\
  $V$   & $-0.379 \pm$ 0.016 & (V-R) & 3.925 $\pm$ 0.017 & 10  & 0.012 & 0.72-1.32 & $0.758 \pm$ 0.032 & 0.591 $\pm$ 0.034 & giants & FG97 \\
  $K$   & $-0.100 \pm$ 0.016 & (J-K) & 3.940 $\pm$ 0.013 & 10  & 0.008 & 0.60-1.06 & $0.200 \pm$ 0.032 & 0.561 $\pm$ 0.026 & giants & FG97 \\
  $V$   & $-0.119 \pm$ 0.007 & (V-K) & 3.914 $\pm$ 0.023 & 13  & 0.032 & 0.52-5.53 & $0.238 \pm$ 0.014 & 0.613 $\pm$ 0.046 & supergiants & FG97\\
  $V$   & $-0.392 \pm$ 0.025 & (V-R) & 3.943 $\pm$ 0.026 & 12  & 0.033 & 0.23-1.78 & $0.784 \pm$ 0.050 & 0.555 $\pm$ 0.052 & supergiants & FG97\\
  $K$   & $-0.101 \pm$ 0.027 & (J-K) & 3.938 $\pm$ 0.023 & 13  & 0.027 & 0.17-1.21 & $0.202 \pm$ 0.054 & 0.565 $\pm$ 0.046 & supergiants & FG97\\
  $V$   & $-0.131 \pm$ 0.003 & (V-K) & 3.947 $\pm$ 0.003 & 10  &       &           & $0.262 \pm$ 0.006 & 0.547 $\pm$ 0.006 & cepheids & FG97, forced common ZP \\
  $V$   & $-0.380 \pm$ 0.003 & (V-R) & 3.947 $\pm$ 0.003 & -   &       &           & $0.760 \pm$ 0.006 & 0.547 $\pm$ 0.006 & cepheids & FG97, forced common ZP \\
  $K$   & $-0.110 \pm$ 0.003 & (J-K) & 3.947 $\pm$ 0.003 & 11  &       &           & $0.220 \pm$ 0.006 & 0.547 $\pm$ 0.006 & cepheids & FG97, forced common ZP \\
  $V$   & $-0.122 \pm$ 0.001 & (V-K) & 3.927 $\pm$ 0.003 &   5 & 0.051 & 1.4-3.7   & $0.244 \pm$ 0.002 & 0.587 $\pm$ 0.006 & (G0-K5) giants & DB93 \\
  $V$   & $-0.101 \pm$ 0.002 & (V-K) & 3.833 $\pm$ 0.008 &   6 & 0.090 & 3.7-6.3   & $0.202 \pm$ 0.004 & 0.775 $\pm$ 0.016 & M-giants & DB93 \\
  $V$   & $-0.139 \pm$ 0.001 & (V-K) & 3.958 $\pm$ 0.005 &  13 & 0.120 & 0.5-4.5   & $0.278 \pm$ 0.002 & 0.525 $\pm$ 0.01 & supergiants & DB93 \\
%  $V$   & $-0.321$           & (V-R) & 3.841             &  39 & 0.033 & 0.80-4.18 & $0.642$           & 0.759            & I/III & BEM78 \\
  $V$   &            & (V-K) &           &  27 & 0.005 & -0.3-4.1 & $0.275 \pm$ 0.001 & 0.518 $\pm$ 0.003  & {\sc iv + v} & Kervella et al. (2004) \\
  $V$   &            & (V-R) &           &  23 & 0.044 &  0.0-1.5 & $0.790 \pm$ 0.008 & 0.522 $\pm$ 0.006  & {\sc iv + v} & Kervella et al. (2004) \\
  $K$   &            & (J-K) &           &  27 & 0.029 &  0.1-0.9 & $0.319 \pm$ 0.007 & 0.509 $\pm$ 0.003  & {\sc iv + v} & Kervella et al. (2004) \\

\hline
\end{tabular}
$^{(1)}$. rms listed in the original publication, which refer to
Eq.~(4) [except entries from Kervella et al. 2004]. In these cases,
multiply by a factor of 2 to compare to the rms values listed in Table~\ref{TAB-RES}.
\label{TAB-PREV}
\end{table*}

\begin{table}
\setlength{\tabcolsep}{1.3mm}
\caption{Colours of normal solar metallicity stars (from Pietrinferni et al. 2004).}
\begin{tabular}{rrrrrrr} \hline

$T_{\rm eff}$ & $\log g$ & $(B-V)$ & $(V-I)$ & $(V-R)$ & $(J-K)$ & $(V-K)$ \\
\hline

 9500 & 2.0 &  -0.043 & 0.022 & 0.003 &  -0.002 & 0.003  \\
 9750 & 1.0 &  -0.022 & 0.060 & 0.026 &  -0.003 & 0.004  \\
10000 & 0.5 &  -0.007 & 0.079 & 0.039 &  -0.003 & 0.002  \\
10250 & 0.0 &   0.007 & 0.100 & 0.050 &   0.000 & 0.004   \\
\hline
\end{tabular}
\label{TAB-COL}
\end{table}

\paragraph*{Acknowledgements}

This research has made use of the SIMBAD database, operated at CDS,
Strasbourg, France. I thank Ren\'e Oudmaijer and Pascal Fouqu\'e for
reading a preliminary version of this paper, and Maurizio Salaris for
discussion.

{}

\end{document}